\RequirePackage{fix-cm}
\let\accentvec\vec
\documentclass[twocolumn]{svjour3}          

     \let\vec\accentvec
     
\usepackage{amsmath}
\smartqed  

\usepackage{examplep}  

\usepackage{graphicx}
\usepackage{cite}
\usepackage{algorithmicx}
\usepackage[ruled]{algorithm}
\usepackage{algpseudocode}
\usepackage{mathtools}
\usepackage{dsfont}

\usepackage{amsfonts}
\usepackage{amssymb}
\usepackage{graphicx}
\usepackage{multirow}
\usepackage{textcomp}
\usepackage{multicol}
\usepackage[usestackEOL]{stackengine}
\usepackage[hyphens]{url}
\usepackage{pict2e}
\usepackage{color}
\usepackage[keeplastbox]{flushend}
\usepackage{enumerate}
\usepackage{float}
\usepackage{mathrsfs}
\usepackage{mathtools}
\usepackage{tikz}

\DeclarePairedDelimiter{\ceil}{\lceil}{\rceil}
\DeclarePairedDelimiter{\floor}{\lfloor}{\rfloor}

\begin{document}

\title{Authenticated Secret Key Generation in Delay Constrained Wireless Systems
}

\author{Miroslav Mitev          \and
        Arsenia Chorti         \and
        Martin Reed            \and
        Leila Musavian
}


\institute{Correspondence: M. Mitev  \at
              School of CSEE, University of Essex, Colchester, UK \\
              \email{mm17217@essex.ac.uk}          
           \and
           A. Chorti \at
              ETIS UMR8051, CY University, ENSEA, CNRS, F-95000, Cergy, France \\
            \email{arsenia.chorti@ensea.fr}
            \and
            M. Reed \at 
            School of CSEE, University of Essex, Colchester, UK \\
              \email{mjreed@essex.ac.uk}
              \and
               L. Musavian \at 
            School of CSEE, University of Essex, Colchester, UK \\
              \email{leila.musavian@essex.ac.uk}
}

\date{Received: date / Accepted: date}

\maketitle
\begin{abstract}

With the emergence of 5G low latency applications, such as haptics and V2X, low complexity and low latency security mechanisms are needed. Promising  lightweight mechanisms include physical unclonable functions (PUF) and secret key generation (SKG) at the physical layer, as considered in this paper. In this framework we propose \textcolor{black}{i) a zero-round-trip-time (0-RTT) resumption authentication protocol combining PUF  and SKG processes; ii) a novel authenticated encryption (AE) using SKG; iii) pipelining of the AE SKG and the encrypted data transfer in order to reduce latency. Implementing the pipelining at PHY, we investigate a \emph{parallel} SKG approach for multi-carrier systems, where a subset of the subcarriers are used for SKG and the rest for data transmission. The optimal solution to this PHY resource allocation problem} is identified under security, power and delay constraints, by formulating the subcarrier scheduling as a subset-sum $0-1$ knapsack optimization. A heuristic algorithm of linear complexity is proposed and shown to incur negligible loss with respect to the optimal dynamic programming solution. All of the proposed mechanisms, have the potential to pave the way for a new breed of latency aware security protocols.

\end{abstract}

\section{Introduction}
\label{sec:intro}
Many standard cryptographic schemes, particularly tho\-se in the realm of public key encryption (PKE), are computationally intensive, incurring considerable overheads and can rapidly drain the battery of power constrained devices \cite{Amitav}, \cite{Aylin}, notably in Internet of things (IoT) applications \cite{Chorti_sensors}. For example, a 3GPP report on the security of ultra reliable low latency communication (URLLC) systems notes that authentication for URLLC is still an open problem \cite{3gppURLLC}.  Additionally, traditional public key generation schemes are not \textit{quantum secure} -- in that when sufficiently capable quantum computers will be available they will be able to break current known PKE schemes -- unless the key sizes increase to impractical lengths.

In the past years, physical layer security (PLS) \cite{A_Chorti_PLS1, A_Chorti_PLS2, A_Chorti_PLS3, A_Chorti_PLS0, Mitev_sub-scheduling} has been studied as a possible alternative to classic, complexity based, cryptography. As an example, signal properties as in \cite{Chorti_FTN}, can be exploited to generate opportunities for confidential data transmission \cite{Chorti_secFTN, A_Chorti_PLS4}. Notably, PLS is explicitly mentioned as a 6G enabling technology in the first white paper on 6G~\cite{6g_white_paper}: ``The strongest security protection may be achieved at the physical layer.''  In this work, we propose to move some of the security core functions down to the physical layer, exploiting both the communication radio channel and the hardware, as unique entropy sources.

Since the wireless channel is reciprocal, time-variant and random in nature, it offers a valid, inherently secure source that may be used in a key agreement protocol between two communicating parties. The principle of secret key generation (SKG) from correlated observations was first studied in \cite{Maurer} and \cite{Csiszar}. A straightforward SKG approach can be built by exploiting the reciprocity of the wireless fading coefficients between two terminals within the channel coherence time\cite{Ye} and this paper builds upon this mechanism. This is pertinent to many forthcoming B5G applications that will require a strong, but nevertheless, lightweight security key agreement; in this direction, PLS may offer such a solution, or, complement existing algorithms. With respect to authentication, physical unclonable functions (PUFs), \textcolor{black}{firstly introduced in \cite{First_silicon_puf} (based on the idea of physical one-way functions~\cite{First_PUF})}, \cite{PUF_Survey} could also enhance authentication and key agreement in demanding scenarios, including (but not limited to) device to device and tactile Internet. We note that others also point to using physical layer security to reduce the resource overhead in URLLC~\cite{Weinand2018}. 

A further advantage of PLS is that it is information-theoretic secure \cite{PLS_Survey}, \textit{i.e.}, it is not open to attack by future quantum computers, and, it requires lower computation costs as will be explored later in this paper. In this work, we will discuss how SKG from shared randomness \cite{Chorti} is a promising alternative to PKE for key agreement. 
\textcolor{black}{However, unauthenticated key generation is vulnerable to man in the middle (MiM) attacks. In this sense, PUFs, 
can be used in \textit{conjunction} with SKG to provide authenticated encryption (AE). }
As summarised in \cite{PUF_Survey} the employment of PUFs can decrease the computational cost and have a high impact on reducing the authentication latency in constrained devices.

\textcolor{black}{In this study we introduce the joint use of PUF authentication and SKG in a zero-round-trip-time (0-RTT) \cite{TLS1_3, 0-RTT_example} approach, allowing to build quick authentication mechanisms with forward security. Further, we develop an AE primitive \cite{AE_Bellare, AE_Krovetz, AE_Das} based on standard SKG schemes. To investigate a fast implementation of the AE SKG we propose a pipelined (\emph{parallel}) scheduling method for optimal resource allocation at the physical layer (PHY) (\textit{i.e.}, by optimal allocation of the subcarriers in 5G resource blocks). }

Next, we extend the analysis to account for statistical delay quality of service (QoS) guarantees, a pertinent scenario in B5G. The support of different QoS guarantee levels is a challenging task. In fact, in time-varying channels, such as in wireless networks, determining the exact delay-bound depending on the users' requirements, is impossible. However, a practical approach, namely the effective capacity\cite{Negi}, can provide statistical QoS guarantees, and, can give delay-bounds with a small violation probability. In our work, we employ the effective capacity as the metric of interest and investigate how the proposed pipelined AE SKG scheme performs in a delay-constrained scenario. 

The system model introduced in this work assumes that a block fading additive white Gaussian noise (BF-AWGN) channel is used with multiple orthogonal subcarriers. In our \emph{parallel} scheme a subset of the subcarriers is used for SKG and the rest for encrypted data transfer. The findings of this paper are supported by numerical results, and the efficiency of the proposed \emph{parallel} scheme is shown to be greater or similar to the efficiency of an alternative approach in which SKG and encrypted data transfer are sequentially performed.

\begin{figure}[t]
\setlength{\unitlength}{0.07in} 
\centering
\begin{picture}(23.5,19.5) 
\linethickness{1pt}
\put(3.5,11.5){\oval(30,12)}
\put(19.5,7.5){\oval(30,12)}
\put(-1,14.5){\Longstack[c]{\large 0-RTT }}
\put(15,2.5){\Longstack[c]{ \large AE SKG }}
\put(-4.5,9.25){\oval(11,6)}
\put(-9.5,8){\Longstack[c]{\scriptsize PUF \\ \scriptsize authentication }}
\put(11.5,9.25){\oval(11,6)}
\put(10,9){\Longstack[c]{\scriptsize SKG }}
\put(27.5,9.25){\oval(11,6)}
\put(24,8){\Longstack[c]{\scriptsize Resource \\ \scriptsize allocation }}
\end{picture}
\caption{\textcolor{black}{Roadmap of contributions.}}
\label{fig:contributions}
\end{figure}
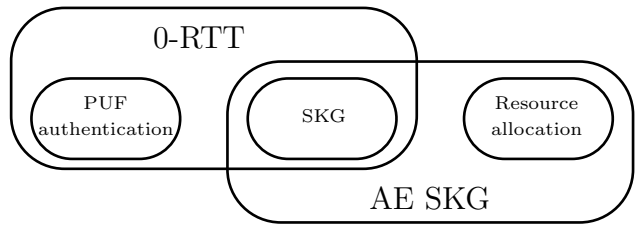

\textcolor{black}{The contributions of this paper are as follows:
\begin{enumerate}
\item  We combine an initial PUF  authentication and SKG for resumption key derivation in a single 0-RTT protocol.
\item We develop an AE SKG scheme.
    \item We propose a fast implementation of the AE SKG based on pipelining of key generation and encrypted data transfer. This \emph{parallel} approach is achieved by allocation of the PHY resources, \textit{i.e.}, by optimal scheduling of the subcarriers in BF-AWGN channels.
    \item We propose a heuristic algorithm of linear complexity that finds the optimal subcarrier allocation with negligible loss in terms of efficiency. 
    \item We numerically compare the efficiency of our \emph{parallel} approach with a \emph{sequential} approach where SKG and data transfer are performed sequentially. This comparison is performed in two delay scenarios:           
    \begin{itemize}
        \item When a relaxed QoS delay constraint is in place;
        \item When a stringent QoS delay constraint is in place.
    \end{itemize}
\end{enumerate}
}
\noindent
\textcolor{black}{A roadmap of the paper's contributions is shown in Fig. \ref{fig:contributions}. }

 The paper is organized as follows: related work is discussed in Section \ref{sec:related_work} followed by a brief summary of the methods used within this paper in Section \ref{sec:methods} and then the general system model is introduced in Section 4. The use of PUF authentication is illustrated in Section \ref{subsec:PUF}, the baseline SKG in Section 4.2; next, in Sections \ref{subsec:key_integrity} and \ref{subsec:0RTT} we present an AE scheme using SKG and a resumption scheme to build a 0-RTT protocol.  Subsequently, we evaluate the optimal power and subcarrier allocation at PHY considering both the long term average rate in Section \ref{sec:hybrid} and the effective rate in Section \ref{sec:eff_capacity}. In Section \ref{sec:results}, the efficiency of the proposed approach is evaluated against that of a sequential approach, while conclusions are presented in Section \ref{sec:conclusions}.


\section{Related Work}
\label{sec:related_work}
\color{black}

This paper assumes the use of PUF-based authentication with SKG. PUFs are hardware entities based on the physically unclonable variations that occur during the production process of silicon. These unique and unpredictable variations allow the extraction of uniformly distributed binary sequences. Due to their unclonability and simplicity, PUFs are seen as lightweight security primitives that can offer alternatives to today's authentication mechanisms. Furthermore, employing PUFs can eliminate the need of non-volatile memory, which reduces cost and complexity~\cite{MutualPufAuth}. Common ways of extracting  secret bit sequences are through measuring delays on wires and gates or observing the power up behavior of a silicon. 

Focusing on that, numerous PUF architectures have been proposed for IoT applications in the literature. A few of these architectures are: arbiter PUF~\cite{Arbiter_PUF}, ring oscillator PUF~\cite{First_silicon_puf}, transient effect ring oscillator PUF~\cite{TERO_PUF}, static random-access memory PUF~\cite{SRAM_PUF},  hardware embedded delay PUF~\cite{HELP_PUF} and more~\cite{PUFs_challenges}. Utilising these basic properties, many PUF-based authentication protocols have been proposed, both for unilateral authentication \cite{ PUF-oneway-auth, PUF_auth-one-two-way} and mutual authentication \cite{PUF_auth-one-two-way, MutualPufAuth, PUF_e-cash_mutual_auth, PUF_mutual_auth1}. A comprehensive survey on lightweight PUF authentication schemes is presented by Delvaux \textit{et al.}~\cite{PUF_auth_survey}.

On the other hand, due to the nature of propagation in a shared free-space, wireless communication remains vulnerable to different types of attacks. Passive attacks such as eavesdropping or traffic analysis can be performed by anyone in the vicinity of the communicating parties; to ensure confidentiality data encryption is vital for communication security. The required keys can be agreed at PHY using SKG. 
In this case, all pilot exchanges need to take place over the coherence time of the channel\footnote{ The coherence time corresponds to the interval during which the multipath properties of wireless channels (channel gains, signal phase, delay) remain stable \cite{Jana1,Rappaport,SKG_vehicular2}. It  is inversely proportional to the Doppler spread, which on the other hand, is a dispersion metric that accounts for the spectral broadening caused by the user's mobility (for more details and derivation please see \cite{Rappaport}).}, during which Alice and Bob can observe highly correlated channel states that can be  used  to generate a shared secret key between them. 
SKG has been implemented and studied for different applications such as vehicular communications \cite{SKG_vehicular1,SKG_vehicular2}, underwater communications \cite{SKG_underwater}, optical fiber \cite{SKG_optical}, visible light communication \cite{SKG_visible_light_communication} and more as summarized in \cite{SKG_implementation_survey}. The key conclusion from these studies is that SKG shows promise as an important alternative to current key agreement schemes. 

Widely used sources of shared randomness used for SKG are the received signal strength (RSS) and the full channel state information (CSI)~\cite{channel_estimation_techniques}. In either case, it is important to build a suitable pre-processing unit to decorrelate the signals in the time / frequency and space domains. 
As an example, some recent works have shown that the widely adopted assumption~\cite{half_wavelength} that a distance equal to \textit{half} of the wavelength (which at $2.4$ GHz is approximately $6$ cm \cite{CSI_immune}) is enough for two channels to  decorrelate, may not hold in reality \cite{Jana1}. Other works show that the mobility can highly increase the entropy of the generated key~\cite{SKG_experiment1, SKG_experiment3} while an important issue with the RSS-based schemes is that they are open to predictable channel attacks\cite{Jana1, RSS_prediction}. These important issues need to be explicitly accounted for in actual implementations, but fall outside the scope of this paper.

\color{black}

\section{Methods}
\label{sec:methods}
\color{black}
The methods used and introduced in this paper rely upon a range of basic primitives, which, in combination provide the full PUF and AE SKG solution. Each of these primitives is introduced below together with a summary of the methods used to analyse and optimise the solution.
\begin{description}
    \item[\textit{Authentication}:]  Before establishing a shared secret key, Alice and Bob must be sure they are communicating with a trusted party. To achieve this we assume the usage of a PLS method, more specifically PUF authentication. As discussed in Section \ref{sec:related_work} by eliminating the need of non-volatile memory the usage of PUFs could greatly reduce the complexity compared to existing authentication alternatives.
    \item[\textit{Secret key generation}:]  To ensure that their communication is private, after authenticating each other Alice and Bob have to encrypt / decrypt the data. For this work we assume the use of symmetric encryption where the same key is used for both operations. In order to obtain a shared key we propose to use SKG which consists of three standard steps: i) advantage distillation; ii) information reconciliation; and, iii) privacy amplification; each of these steps is explained in more detail in Section \ref{subsec:system model}.   
    \item[\textit{Re-authentication}:]  We present a re-authentication approach that exploits the use of resumption secrets as used in 0-RTT protocols. Instead of performing full authentication before sending data encrypted with a new key, we propose a new method which allows Alice (Bob) to authenticate subsequent keys using a lightweight scheme anchored by the initial authentication process. 
    \item[\textit{Authenticated encryption SKG}:]  To eliminate the possibility of tampering attacks we build on the SKG process to introduce a new AE SKG method. AE  can simultaneously guarantee confidentiality and message integrity. In our AE SKG method, side information and encrypted data transfer are pipelined.
    \item[\textit{Pipelined transmission}:] In our proposal, the key generation is pipelined with the encrypted data transfer, \textit{i.e.}, side information and data encrypted with the key that corresponds to the side information are transmitted over the same 5G resource block(s). 
    \item[\textit{Joint PHY/MAC delay analysis}:] To analyze the system under statistical QoS delay constraints we use 
    the theory of \textit{effective capacity} \cite{Negi} and analyse the sche\-me's \textit{effective rate}. 
    \item[\textit{Optimization methods}:]  Finally, to optimize the pipelined transmission, we take into consideration practical wireless aspects such as the impact of imperfect CSI measurements and formulate two optimization problems to find the optimal resource allocation for Alice and Bob. 
    To solve these problems we employ tools such as combinatorial optimization, dynamic programming, order statistics and convex optimization. 
\end{description}
\color{black}

\subsection{Threat Model}
\color{black}
In this paper we assume a commonly used adversarial model  with an active man-in-the-middle attacker (Eve) and a pair of legitimate users (Alice and Bob). For simplicity, we assume a rich Rayleigh multipath environment where the adversary is more than a few wavelengths away from each of the legitimate parties. This forms the basis of our hypothesis that the measurements  of Alice and Bob are uncorrelated to the Eve's measurements. 

\subsection*{Notation}

Random variables are denoted in italic font, e.g., $x$, vectors and matrices are denoted with lower and upper case bold characters, e.g., $\mathbf{x}$ and $\mathbf{X}$, respectively. Functions are printed in a fixed-width teletype font, e.g., $\verb"F"$. All sets of vectors are given with calligraphic font $\mathcal{X}$ and the elements within a set are given in curly brackets e.g. $\{ \mathbf{x}, \mathbf{y}\}$, the cardinality of a vector or set is defined by vertical lines e.g., $|\mathbf{x}|$ or $|\mathcal{X}|$ . Concatenation and bit-wise XORing are represented as $[ \mathbf{x} || \mathbf{y}] $ and $\mathbf{x} \oplus \mathbf{y}$, respectively. We use $H$ to denote entropy, $I$ mutual information,  $\mathbb{E}$ expectation and $\mathbb{C}$ the set of complex numbers.
\color{black}
\section{ Node Authentication Using PUFs and SKG }
\label{sec:three}
\textcolor{black}{In this Section we present a joint physical layer SKG and PUF authentication scheme. To the best of our' knowledge this is the first work that proposes the utilization of the two schemes in conjunction. As discussed in Section \ref{sec:related_work}, many PUF authentication protocols have been proposed in the literature, with even a few commercially available~\cite{PUF_commercial1, PUF_commercial2}. We do not look into developing a new PUF architecture or a new PUF authentication protocol, instead, we look at combining existing PUF mechanisms with SKG. In addition, we develop an AE scheme that can prevent tampering attacks. To further develop our hybrid crypto-system we propose a resumption type of authentication protocol, inspired by the 0-RTT authentication mode in the transport layer security (TLS) 1.3 protocol. The resumption protocol is important as it significantly reduces the use of the PUF to the initial authentication, thus, overcoming the limitation of a PUFs' challenge response space~\cite{PUFs_challenges, PUF_limited_CRP}.}

\subsection{Node Authentication Using PUFs}
\label{subsec:PUF}

\textcolor{black}{As discussed in Section \ref{sec:methods}}, for security against  MiM attacks, the SKG needs to be protected through authentication. While existing techniques, such as the extensible authentication protocol-transport layer security (EAP-TLS), could be used as the authentication mechanism,  these are computationally intensive and can lead to significant latency \textcolor{black}{\cite{comp_intensive_eap-tls1, comp_intensive_eap-tls2}}.

This leads to the motivation to seek lightweight authentication mechanisms that can be used in conjunction with SKG. Such a mechanism that is achieving note within the research community uses a PUF. The concept of a PUF was first introduced in \textcolor{black}{\cite{First_silicon_puf}}, its idea is to utilize the fact that every integrated circuit differs to others due to manufacturing variability \cite{PUF, PUF_Suh} and cannot be cloned \cite{PUF_unclonable}. Having these characteristics a PUF can be used in a challenge -- response scheme, where a challenge can refer to a delay at a specific gate, power-on state, etc. 

A typical PUF-based authentication protocol consists of two main phases, namely \textit{enrolment phase} and \textit{authentication phase} \cite{PUF_Chat, PUF_Location, PUF_Wifi, PUF_Braeken, PUF_TS}. During the \textit{enrolment pha\-se} each node runs a set of challenges on its PUF and characterizes the variance of the measurement noise in order to generate side information. Next, a verifier creates and stores a database of all challenge-response pairs (CRPs) for each node's PUF within its network. A CRP pair in essence consists of an  authentication key and related side information. Within the database, each CRP is associated with the ID of the corresponding node.

Later, during the \textit{authentication phase} a node sends its ID to the verifier requesting to start a communication. Receiving the request, the verifier checks if the received ID exists in its database. If it does, the verifier chooses a random challenge that corresponds to this ID and sends it to the node. The node computes the response by running the challenge on its PUF and sends it to the verifier. However, the PUF measurements at the node are never exactly the same due to measurement noise, therefore, the verifier uses the new PUF measurement and the side information stored during the enrollment to re-generate the authentication key. Finally, the verifier compares the re-generated key to the one in the CRP and if they are identical the authentication of the node is successful. In order to prevent replay attacks, once used, a CRP is deleted from the verifier database.

In summary, the motivation for using a PUF authentication scheme in conjunction with SKG is to exclude all of the computationally intensive operations required by EAP-TLS, which use modulo arithmetic in large fields. Measurements performed on current public key operations within EAP-TLS on common devices (such as IoT) give  average authentication and key generation times of approximately 160 ms in static environments and this can reach up to 336 ms in high mobility conditions \cite{EAP_Ahmad}. 

On the other hand, PUF authentication protocols have very low computational overhead and require overall authentication times that can be less than 10 ms \cite{PUF_Gope, PUF_Location}. Furthermore, our key generation scheme, proposed in Section \ref{subsec:system model}, requires just a hashing operation and (syndrome) decoding. Hashing mechanisms such as SHA256 performed on an IoT device require less than 0.3ms \cite{PUF_Gope, Hash_Ometov}. Regarding the decoding, if we assume the usage of standard LDPC or BCH error correcting mechanisms, even in the worst-case scenario with calculations carried out as software operations, the computation is trivial compared to the hashing and requires less computational overhead \cite{BCH_Cho}.

\subsection{SKG procedure}
\label{subsec:system model}

\begin{figure}
\tikzset{>=latex}
\begin{tikzpicture}
\begin{scope}[xshift=2cm]
\draw[rounded corners=15pt,line width=0.71pt] (0.5,0)--(3,0)--(3,9)--(0,9)--(0,0)--(.5,0);
\node at (1.5,9.5) {\large Alice};
\draw[thick, ->] (3,8.4) -- (5.2,8.4);
\draw[thick, ->] (5.2,8.2)--(3,8.2);
\node at (3.4,8.6) {\scriptsize Pilots};
\node at (4.7,8) {\scriptsize Pilots};
\node at (1.5,8) {$\mathbf{x}_A$};
\draw[thick, ->] (1.5,7.8)--(1.5,7);
\node[draw, rounded corners=2pt,line width=1pt, text width=1.6cm, align=center] at (1.5,6) {\Longstack[c]{\scriptsize Quantizer}};
\node[draw, rounded corners=2pt,line width=1pt, text width=1.6cm, align=center] at (1.5,5.3) {\Longstack[c]{\scriptsize Slepian Wolf \\ \scriptsize decoder }};
\draw[rounded corners=5pt,line width=1pt,densely dashed] (1,4.7)--(2.7,4.7)--(2.7,7)--(0.3,7)--(0.3,4.7)--(1,4.7);
\node at (1.5,6.6) {\Longstack[c]{\scriptsize Information \\ \scriptsize reconciliation}};
\draw[thick, ->] (2.7,5.85)--(5.45,5.85);
\node at (4.05,6.05) {$\mathbf{s}_A$};
\draw[thick, ->] (1.5,4.7)--(1.5,3.9);
\node at (1.75,4.3) {$\mathbf{r}_A$};
\node[draw, rounded corners=2pt,line width=1pt, text width=1.6cm, align=center] at (1.5,3.5) {\Longstack[c]{\scriptsize Privacy \\ \scriptsize amplification }};
\node at (1.5,2.9) {\scriptsize Key \normalsize  $\mathbf{k}$};
\draw[thick, ->] (1.5,2)--(6.6,2);
\node at (4.05,2) {\Longstack[c]{\scriptsize Encrypted data  \\ \scriptsize with key \normalsize $\mathbf{k}$}};
\end{scope}

\begin{scope}[xshift=7.2cm]
\draw[rounded corners=15pt,line width=1pt] (0.5,0)--(3,0)--(3,9)--(0,9)--(0,0)--(.5,0);
\node at (1.5,9.5) {\large Bob};
\node at (1.5,8) {$\mathbf{x}_B$};
\draw[thick, ->] (1.5,7.8)--(1.5,7);
\node[draw, rounded corners=2pt,line width=1pt, text width=1.6cm, align=center] at (1.5,6) {\Longstack[c]{\scriptsize Quantizer}};
\node[draw, rounded corners=2pt,line width=1pt, text width=1.6cm, align=center] at (1.5,5.3) {\Longstack[c]{\scriptsize Slepian Wolf \\ \scriptsize decoder }};
\draw[rounded corners=5pt,line width=1pt,densely dashed] (1,4.7)--(2.7,4.7)--(2.7,7)--(0.3,7)--(0.3,4.7)--(1,4.7);
\node at (1.5,6.6) {\Longstack[c]{\scriptsize Information \\ \scriptsize reconciliation}};
\draw[thick, ->] (1.5,4.7)--(1.5,3.9);
\node at (1.75,4.3) {$\mathbf{r}_A$};
\node[draw, rounded corners=2pt,line width=1pt, text width=1.6cm, align=center] at (1.5,3.5) {\Longstack[c]{\scriptsize Privacy \\ \scriptsize amplification }};
\node at (1.5,2.9) {\scriptsize Key \normalsize  $\mathbf{k}$};
\end{scope}
\end{tikzpicture}
\caption{\textcolor{black}{Secret key generation between Alice and Bob. }} \label{fig:skg-procedure}
\end{figure}
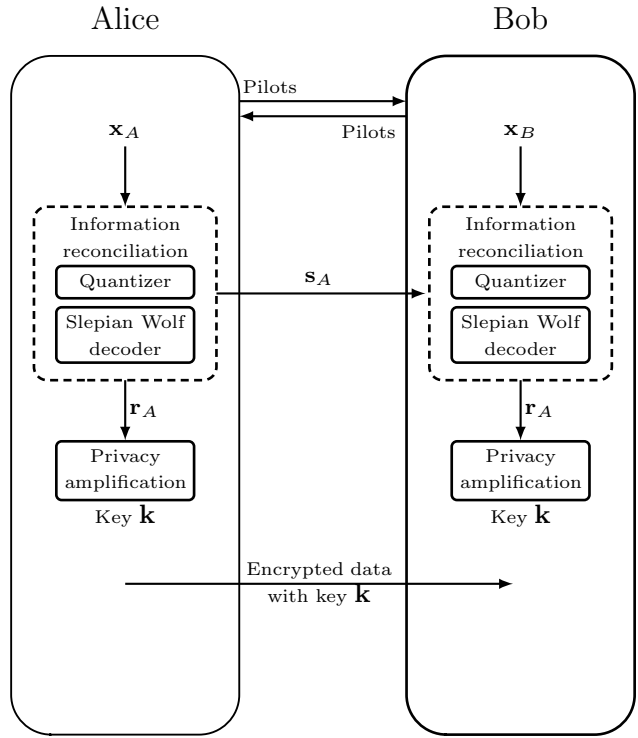

The SKG system model is shown in Fig. \textcolor{red}{\ref{fig:skg-procedure}}. This assumes that two legitimate parties, Alice and Bob, wish to establish a symmetric secret key using the wireless fading coefficients as a source of shared randomness.  Throughout our work a rich Rayleigh multipath environment is assumed, such that the fading coefficients rapidly decorrelate over short distances \cite{Ye}. Furthermore, Alice and Bob communicate over a BF-AWGN channel that comprises $N$ orthogonal subcarriers. The fading coefficients $\mathbf{h} = [{h}_1, \dots , {h}_N]$, are assumed to be independent and identically distributed (i.i.d), complex circularly symmetric zero-mean Gaussian random variables ${h_j}\sim \mathcal{CN}(0, \sigma^2), j=1,\ldots,N$. Although in actual multicarrier systems neighbouring subcarriers will typically experience correlated fading, in the present work this effect is neglected as its impact on SKG has been treated in numerous contributions in the past \cite{Chen_Jensen, Zhang, Zhang_He} and will not enhance the problem formulation in the following sections.

The SKG procedure encompasses three phases: \textit{advantage distillation}, \textit{information reconciliation}, and \textit{privacy amplification} \cite{Maurer}, \cite{Csiszar} as described below:

\textit{1) Advantage distillation}: This phase takes place \textcolor{black}{ during the coherence time of the channel}. The legitimate nodes sequentially exchange constant probe signals with power $P$ on all subcarriers\footnote{An explanation of the optimality of this choice under different attack scenarios is discussed in \cite{Chorti}.}, to obtain estimates of their reciprocal CSI. We note in passing that the pilot exchange phase can be made robust with respect to injection type of attacks (that fall in the general category of MiM) as analyzed in \cite{Chorti, Globecom_2019_MiM}. Commonly, the received signal strength (RSS) has been used as the source of shared randomness for generating the shared key, but it is possible to use the full CSI \cite{Saiki15}. At the end of this phase, Alice and Bob obtain observation vectors $\mathbf{x}_{A}=[{x}_{A,1} , \dots , {x}_{A,N}],  \mathbf{x}_{B}=[{x}_{B,1}, \dots, {x}_{B,N}]$, respectively, so that:
\begin{align}
&\mathbf{x}_{A}=\sqrt{P}\mathbf{h} + \mathbf{z}_{A}, \label{eq:XAj}\\
&\mathbf{x}_{B}=\sqrt{P}\mathbf{h} + \mathbf{z}_{B}, \label{eq:XBj}
\end{align}
where  $\mathbf{z}_{A}$ and $\mathbf{z}_{B}$  denote zero-mean, unit variance circu\-larly symmetric complex AWGN random vectors, such that
\begin{math}
(\mathbf{z}_{A}, \mathbf{z}_{B}) \sim \mathcal{CN} (\mathbf{0}, \mathbf{I}_{2N}).
\end{math}
\color{black}On the other hand, Eve observes $\mathbf{x}_{E}=[{x}_{E,1}, \dots , {x}_{E,N}]$ with:
\begin{equation}
    \mathbf{x}_{E}=\sqrt{P}\mathbf{h}_{E} + \mathbf{z}_{E}.
\end{equation}
Due to the rich Rayleigh multipath environment, Eve's channel measurement  $\mathbf{h}_{E}$ is assumed uncorrelated to $\mathbf{h}$ and $\mathbf{z}_{E}$ denotes a zero-mean, unit variance circularly symmetric complex AWGN random vector $\mathbf{z}_{E}  \sim \mathcal{CN}(\mathbf{0}, \mathbf{I}_{N})$. 
 \color{black}
 
\textit{2) Information reconciliation}:  \textcolor{black}{At the beginning of this phase the observations ${x}_{A,j}, {x}_{B,j}$ are quantized to binary vectors\footnote{Note that each observation can generate a multi-bit vector at the output of the quantizer.} $\mathbf{r}_{A,j}$, $\mathbf{r}_{B,j}$ $j=1,\ldots,N$\cite{wang2011fast, Secret_bits_after_IR, Secret_bits_after_IR2}, so that Alice and Bob distill $\mathbf{r}_{A} =[\mathbf{r}_{A,1}|| \dots || \mathbf{r}_{A,N}]$ and $\mathbf{r}_{B}=[\mathbf{r}_{B,1}|| \dots || \mathbf{r}_{B,N}]$, respectively.} Due to the presence of noise, $\mathbf{r}_{A}$ and  $\mathbf{r}_{B}$ will differ. To reconcile  discrepancies in the quantizer local outputs, side information needs to be exchanged via a public channel. Using the principles of Slepian Wolf decoding, the distilled binary vectors can  be expressed as
\textcolor{black}{
\begin{eqnarray}
\mathbf{r}_{A}=\mathbf{d}+\mathbf{e}_{A},\\
\mathbf{r}_{B}=\mathbf{d}+\mathbf{e}_{B},
\end{eqnarray}
where $\mathbf{e}_{A}, \mathbf{e}_{B}$ are error vectors that represent the distance from the common observed (codeword) vector $\mathbf{d}$ at Alice and Bob, respectively.}

Numerous practical information reconciliation approaches using standard forward error correction codes (e.g., LDPC, BCH, etc.,) have been proposed \cite{Ye}, \cite{Saiki15}. As an example, if a block encoder 
is used, then the error vectors can be recovered from the 
syndromes $\mathbf{s}_{A}$ and 
  $\mathbf{s}_{B}$ 
of $\mathbf{r}_{A}$ and $\mathbf{r}_{B}$, respectively. 
Alice transmits her corresponding syndrome to Bob so that 
he can reconcile $\mathbf{r}_{B}$ to $\mathbf{r}_{A}$. 
\textcolor{black}{  It has been shown that the length of the the syndrome $|\mathbf{s}_A|$ is lower bounded by $|\mathbf{s}_A| \geq  H(\mathbf{x}_A|\mathbf{x}_B)=H(\mathbf{x}_A, \mathbf{x}_B)-H(\mathbf{x}_B)$~\cite{Csiszar}. This has been numerically evaluated for different scenarios and coding techniques \cite{Secret_bits_after_IR, Evaluation_leakage1, Evaluation_leakage2, Evaluation_leakage3}.} \textcolor{black}{Following that, the achievable SKG rate is upper bounded by $I(\mathbf{x}_A;\mathbf{x}_B|\mathbf{x}_E)$}.

\textit{3) Privacy amplification}: 
\textcolor{black}{ The secret key is generated by passing $\mathbf{r}_{A}$ through a one-way collision resistant \textit{compression} function i.e., by hashing. }
Note that this final step of privacy amplification,  is executed locally without any further information exchange. 
\textcolor{black}{ The need for privacy amplification arises in order to suppress the entropy revealed due to the public transmission of the syndrome $\mathbf{s}_A$. 
Privacy amplification produces a key of length strictly shorter than $|\mathbf{r}_A|$, at least by $|\mathbf{s}_A|$. At the same time, the goal is for the key to be uniform, i.e., to have maximum entropy.  In brief, privacy amplification \textit{reduces the overall output entropy} while at the same time \textit{increases the entropy per bit} -- compared to the input}.

\textcolor{black}{The privacy amplification is typically performed by applying either cryptographic  hash functions such as those built using the Merkle-Damgard construction, or universal hash functions and has been proven to be secure, in an information theoretic sense, through the leftover hash lemma \cite{LHL-original}. 
As an example \cite{Jana1, jana-similar} use a 2-universal hash family to achieve privacy amplification. }
\textcolor{black}{Summarizing, the maximum key size after privacy amplification is:
\begin{equation}
    |\mathbf{k}| \leq H(\mathbf{x}_A) - I(\mathbf{x}_A;\mathbf{x_E})-H(\mathbf{x}_A|\mathbf{x}_B)-r_0,
\end{equation}
where $H(\mathbf{x}_A)$ represents the entropy of the measurement, $I(\mathbf{x}_A;\mathbf{x_E})$ represents the mutual information between Alice's and Eve's observations, $H(\mathbf{x}_A|\mathbf{x}_B)$ represents the entropy revealed during information reconciliation and $r_0>0$ is an extra security parameter that ensures uncertainty on the key at Eve's side. For details and estimation of these parameters in a practical scenario please see \cite{Privacy_amplification_key-size}. }


As shown in this section the SKG procedure requires only a few simple operations such as quantization, syndrome calculation and hashing.  
In future work we will examine the real possibilities of implementing such a mechanism in practical systems.

\subsection{AE Using SKG}
\label{subsec:key_integrity}

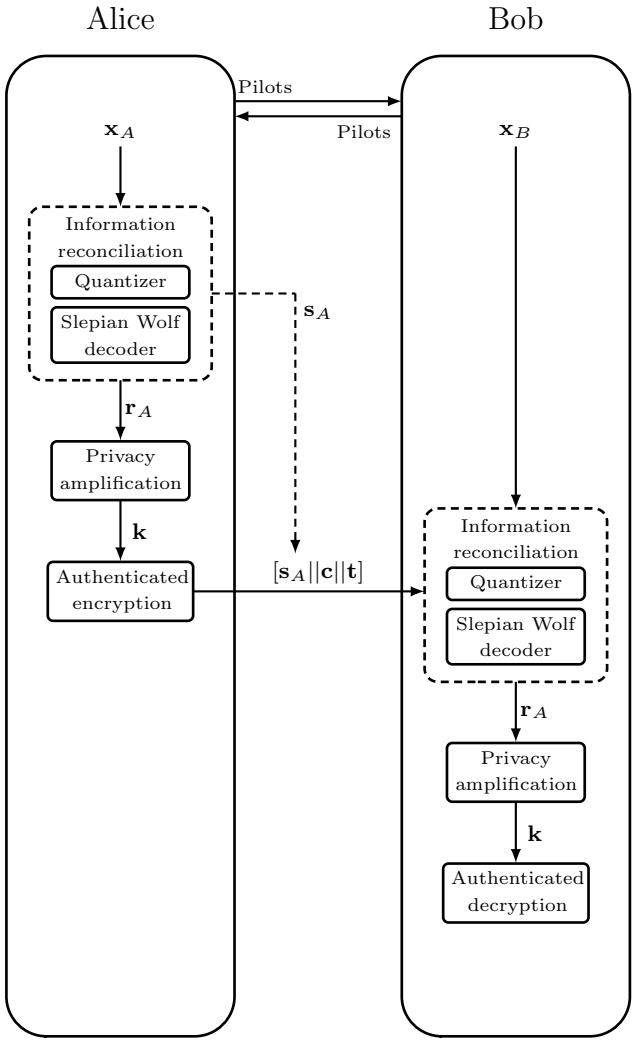
\begin{figure}
\tikzset{>=latex}
\begin{tikzpicture}
\begin{scope}[xshift=2cm]
\draw[rounded corners=15pt,line width=1pt] (0.5,0)--(3,0)--(3,13)--(0,13)--(0,0)--(.5,0);
\node at (1.5,13.5) {\large Alice};
\draw[thick, ->] (3,12.4) -- (5.2,12.4);
\draw[thick, ->] (5.2,12.2)--(3,12.2);
\node at (3.4,12.6) {\scriptsize Pilots};
\node at (4.7,12) {\scriptsize Pilots};
\node at (1.5,12) {$\mathbf{x}_A$};
\draw[thick, ->] (1.5,11.8)--(1.5,11);
\node[draw, rounded corners=2pt,line width=1pt, text width=1.6cm, align=center] at (1.5,10) {\Longstack[c]{\scriptsize Quantizer}};
\node[draw, rounded corners=2pt,line width=1pt, text width=1.6cm, align=center] at (1.5,9.3) {\Longstack[c]{\scriptsize Slepian Wolf \\ \scriptsize decoder }};
\draw[rounded corners=5pt,line width=1pt,densely dashed] (1,8.7)--(2.7,8.7)--(2.7,11)--(0.3,11)--(0.3,8.7)--(1,8.7);
\node at (1.5,10.6) {\Longstack[c]{\scriptsize Information \\ \scriptsize reconciliation}};
 \draw[thick,densely dashed] (2.7,9.85)--(3.8,9.85);
  \draw[thick,densely dashed, ->] (3.8,9.85)--(3.8,6.4);
\node at (4.1,9.6) {$\mathbf{s}_A$};
\draw[thick, ->] (1.5,8.7)--(1.5,7.9);
\node at (1.75,8.3) {$\mathbf{r}_A$};
\node[draw, rounded corners=2pt,line width=1pt, text width=1.6cm, align=center] at (1.5,7.5) {\Longstack[c]{\scriptsize Privacy \\ \scriptsize amplification }};
\draw[thick, ->] (1.5,7.1)--(1.5,6.3);
\node at (1.75,6.7) {$\mathbf{k}$};
\node[draw,  rounded corners=2pt,line width=1pt, text width=1.7cm] at (1.5,5.9) {\Longstack{\scriptsize Authenticated \\  \scriptsize encryption }};
\draw[thick, ->] (2.45,5.9)--(5.5,5.9);
\node[align=left] at (4.1,6.15) {\Longstack[c]{ $[\mathbf{s}_A||\mathbf{c}||\mathbf{t}]$ }};
\end{scope}

\begin{scope}[xshift=7.2cm]
\draw[rounded corners=15pt,line width=1pt] (0.5,0)--(3,0)--(3,13)--(0,13)--(0,0)--(.5,0);
\node at (1.5,13.5) {\large Bob};
\node at (1.5,12) {$\mathbf{x}_B$};
\draw[thick, ->] (1.5,11.8)--(1.5,7);
\node[draw, rounded corners=2pt,line width=1pt, text width=1.6cm, align=center] at (1.5,6) {\Longstack[c]{\scriptsize Quantizer}};
\node[draw, rounded corners=2pt,line width=1pt, text width=1.6cm, align=center] at (1.5,5.3) {\Longstack[c]{\scriptsize Slepian Wolf \\ \scriptsize decoder }};
\draw[rounded corners=5pt,line width=1pt,densely dashed] (1,4.7)--(2.7,4.7)--(2.7,7)--(0.3,7)--(0.3,4.7)--(1,4.7);
\node at (1.5,6.6) {\Longstack[c]{\scriptsize Information \\ \scriptsize reconciliation}};
\draw[thick, ->] (1.5,4.7)--(1.5,3.9);
\node at (1.75,4.3) {$\mathbf{r}_A$};
\node[draw, rounded corners=2pt,line width=1pt, text width=1.6cm, align=center] at (1.5,3.5) {\Longstack[c]{\scriptsize Privacy \\ \scriptsize amplification }};
\draw[thick, ->] (1.5,3.1)--(1.5,2.3);
\node at (1.75,2.7) {$\mathbf{k}$};
\node[draw,  rounded corners=2pt,line width=1pt, text width=1.7cm] at (1.5,1.9) {\Longstack{\scriptsize Authenticated \\  \scriptsize decryption }};
\end{scope}
\end{tikzpicture}
\caption{\textcolor{black}{Pipelined SKG and encrypted data transfer between Alice and Bob.} } \label{fig:Our_SKG_process}
\end{figure}

To develop a hybrid cryptosystem that can withstand tampering attacks, SKG can be introduced in standard AE schemes in conjunction with standard block ciphers in counter mode (to reduce latency), e.g., AES GCM. As a sketch of such a primitive, let us assume a system with three parties: Alice who wishes to transmit a secret message $\mathbf{m}$ \textcolor{black}{with size $|\mathbf{m}|$}, to Bob with confidentiality and integrity, and Eve, that can act as a passive and active attacker. The following algorithms are employed: 
\begin{itemize}
    \item The SKG scheme denoted by $\verb"G": \mathbb{C}\rightarrow \mathcal{K} \times \mathcal{S}$, accepting as input the fading coefficients (modelled as complex numbers),  and generating as outputs binary vectors $\mathbf{k}$ and $\mathbf{s}_A$ in the key and syndrome spaces, of sizes \textcolor{black}{$|\mathbf{k}|$ and $|\mathbf{s}_A|$}, respectively, 
    \begin{equation}
        \verb"G"(\mathbf{h})= \left(\mathbf{k}, \mathbf{s}_{A}\right),
    \end{equation}
  where $\mathbf{k}\in\mathcal{K} $ denotes the key obtained from $\mathbf{h}$ after privacy amplification and $\mathbf{s}_A$ 
  is 
   Alice's syndrome.
\item A symmetric encryption algorithm, e.g., AES GCM, denoted by $\verb"Es": \mathcal{K}\times \mathcal{M} \rightarrow \mathcal{C_T} $ where $\mathcal{C_T}$ denotes the ciphertext space with corresponding decryption $\verb"Ds": \mathcal{K}\times \mathcal{C_T} \rightarrow \mathcal{M}$, such that 
\begin{eqnarray}
    \verb"Es"(\mathbf{k}, \mathbf{m})=\mathbf{c},\\
    \verb"Ds"(\mathbf{k}, \mathbf{c})=\mathbf{m},
\end{eqnarray}
for $\mathbf{m}\in \mathcal{M}$, $\mathbf{c}\in\mathcal{C_T}$.

\item A pair of message authentication code (MAC) algorithms, e.g., in HMAC mode, denoted by $\verb"Sign": \mathcal{K}\times \mathcal{M}\rightarrow \mathcal{T}$, with a corresponding verification algorithm $\verb"Ver": \mathcal{K}\times \mathcal{M} \times \mathcal{T} \rightarrow (yes, no)$, such that 
\begin{eqnarray}
&&\verb"Sign" (\mathbf{k}, \mathbf{m})=\mathbf{t},\\
&&\verb"Ver" (\mathbf{k}, \mathbf{m}, \mathbf{t})=\left\{\begin{array}{ll}
\it{yes}, & \text{if integrity verified}\\
\it{no}, & \text{if integrity not verified}
\end{array}\right. 
\end{eqnarray}
\end{itemize}

A hybrid crypto-PLS system for AE SKG can be built as follows:
\begin{enumerate}
    \item The SKG procedure is launched between Alice and Bob generating a key and a syndrome $\verb"G"(\mathbf{h})\!=\!(\mathbf{k},\mathbf{s}_A)$. 
    \item Alice breaks her key into two parts $\mathbf{k}=\{\mathbf{k}_e, \mathbf{k}_i\}$  and uses the first to encrypt the message as  $\mathbf{c}=\verb"Es"(\mathbf{k}_e, \mathbf{m})$. Subsequently, using the second part of the key she signs the ciphertext using the signing algorithm $\mathbf{t}=\verb"Sign"(\mathbf{k}_i, \mathbf{c})$ and transmits to Bob the extended ciphertext $\left[\mathbf{s}_A\| \mathbf{c}\| \mathbf{t}\right]$, as it is depicted in Fig. \ref{fig:Our_SKG_process}. %
    
    \item Bob checks first the integrity of the received ciphertext as follows: from $\mathbf{s}_A$ and his own observation he evaluates $\mathbf{k}=\{\mathbf{k}_e, \mathbf{k}_i\}$ and computes $\verb"Ver"(\mathbf{k}_i, \mathbf{c}, \mathbf{t})$. The integrity test will fail if any part of the extended ciphertext was modified, including the syndrome (that is sent as plaintext); for example, if the syndrome was modified during the transmission, then Bob would not have evaluated the correct key and the integrity test would have failed. 
    \item If the integrity test is successful then Bob decrypts $\mathbf{m}\verb"=Ds"(\mathbf{k}_e, \mathbf{c})$.
\end{enumerate}

\subsection{Resumption Protocol}
\label{subsec:0RTT}

In Section \ref{subsec:PUF} we discussed that using PUF authentication can greatly reduce the computational overhead of a system. Authentication of new keys is required at the start of communication and at each key renegotiation. However, the number of challenges that can be applied to a single PUF is limited. Due to that we present a solution that is inspired by the 0-RTT authentication mode introduced in the 1.3 version of the TLS \cite{TLS1_3}.  The use of 0-RTT obviates the need of performing a challenge for every re-authentication through the use of a resumption secret $\mathbf{r}_s$, thus reducing latency. Another strong motivation for using this mechanism is that it is forward secure in the scenario we are using here \cite{0-RTT_example}. We first briefly describe the TLS 0-RTT mechanism before describing a similarly inspired 0-RTT mechanism applied to the information reconciliation phase of our SKG mechanism.

The TLS 1.3 $0-$RTT handshake works as follows: In the very first connection between client and server a regular TLS handshake is used. During this step the server sends to the client a look-up identifier $\mathbf{k}_l$ for a corresponding entry in session caches or it sends a session ticket. Then both parties derive a resumption secret $\mathbf{r}_s$ using their shared key and the parameters of the session. Finally, the client stores the resumption secret $\mathbf{r}_s$ and uses it when reconnecting to the same server which also retrieves it during the re-connection.  

If session tickets are used the server encrypts the resumption secret using long-term symmetric encryption key, called a session ticket encryption key (STEK), resulting in a session ticket. The session ticket is then stored by the client and included in subsequent connections, allowing the server to retrieve the resumption secret. Using this approach the same STEK is used for many sessions and clients. On one hand, this property highly reduces the required storage of the server, however, on the other hand, it makes it vulnerable to replay attacks and not forward secure. Due to these vulnerabilities, in this work we focus on the session cache mechanism described next.

When using session caches the server stores all resumption secrets and issues a unique look-up identifier $\mathbf{k}_l$ for each client. When a client tries to reconnect to that server it includes its look-up identifier $\mathbf{k}_l$ in the 0-RTT message, which allows the server to retrieve the resumption secret $\mathbf{r}_s$. Storing a unique resumption secret $\mathbf{r}_s$ for each client requires server storage for each client but it provides forward security and resilience against replay attacks, when combined with a key generation mechanisms such as Diffie Hellman (or the SKG used in this paper) which are important goals for security protocols~\cite{0-RTT_example}. In our physical layer 0-RTT, given that a node identifier state would be required for link-layer purposes, the session cache places little comparative load and thus is the mechanism proposed here for (re-)authentication.

The physical layer resumption protocol modifies the information reconciliation phase of Section 3.1 following initial authentication to provide a re-authentication mechanism between Alice and Bob. At the first establishment of communication we assume initial authentication is established, such as the mechanism shown in Section 4.1. During that Alice sends to Bob a look-up identifier $\mathbf{k}_l$. Then, both derive a resumption secret $\mathbf{r}_s$ that is identified by $\mathbf{k}_l$. Note,  $\mathbf{r}_s$ and the session key have the same length \textcolor{black}{$|\mathbf{k}|$}. Then referring to the notation and steps in Section 4.1-4.3:
 
\begin{enumerate}
    \item  Advantage distillation phase is carried out as before (See section \ref{subsec:system model}), where both parties obtain channel observations and obtain the vectors $\mathbf{r}_{A}$ and $\mathbf{r}_{B}$.
    
    \item During the information reconciliation phase both Alice and Bob exclusive-or the resumption secret $\mathbf{r}_s$ with their observations $\mathbf{r}_{A}$ and $\mathbf{r}_{B}$, obtaining syndromes $\mathbf{s}'_A$ and $\mathbf{s}'_B$ with which both parties can carry out reconciliation to obtain the same shared value which is now $\mathbf{r}_A \oplus \mathbf{r}_s$. 
    
    \item The privacy amplification step in Section 4.2 is carried out as before, but now the hashing takes place on $\mathbf{r}_A \oplus \mathbf{r}_s$ to produce the final shared key $\mathbf{k}'$ that is a result of both the shared wireless randomness and the resumption secret.

\end{enumerate} 

Note that the key $\mathbf{k'}$ can only be obtained if both the physical layer generated key and the resumption key are valid and this method can be shown to be forward secure~\cite{0-RTT_example}.

\section{Pipelined SKG and Encrypted Data Transfer} \label{sec:hybrid}

\textcolor{black}{As explained in Section \ref{sec:three}, if Alice and Bob follow the standard sequential SKG process they can exchange encrypted data only  after both of them have distilled the key at the end of the privacy amplification step. In this Section, we propose a method to pipeline the SKG and encrypted data transfer. Alice can unilaterally extract the secret key from her observation and use it to encrypt data transmitted  in the same ``extended'' message that contains the side information (see Fig. \ref{fig:Our_SKG_process}). Subsequently, using the side information, Bob can distill the same key $\mathbf{k}$ and decrypt the received data in one single step.}

We have discussed in Section \ref{subsec:system model} how Alice and Bob can distill secret keys from estimates of the fading coefficients in their wireless link and in Section \ref{subsec:key_integrity} how these can be used to develop an AE SKG primitive. At the same time CSI estimates are prerequisite in order to optimally allocate power across the subcarriers and achieve high data rates\footnote{As an example, despite the extra overhead, in URLLC systems advanced CSI estimation techniques are employed in order to be able to satisfy the strict reliability requirements.}. As a result, a question that naturally arises is whether the CSI estimates (obtained at the end of the pilot exchange phase), should be used towards the generation of secret keys or towards the reliable data transfer, and, furthermore, whether the SKG and the data transfer can be inter-woven using the AE SKG principle. 

In this paper, we are interested in answering this question and shed light into whether following the exchange of pilots Alice should transmit reconciliation information on all subcarriers, so that she and Bob can generate (potentially) a long sequence of key bits, or, alternatively, perform information reconciliation only over a subset of the subcarriers and transmit encrypted data over the rest, exploiting the idea of the AE SKG primitive. Note here that the data can be already encrypted with the key generated at Alice, the sender of the side information, so that the proposed pipelining does not require storing keys for future use. We will call the former approach a \emph{sequential} scheme, while we will refer to the latter as a \emph{parallel} scheme. The two will be compared in terms of their efficiency with respect to the achievable data rates.

A simplified version of this problem, where the reconciliation rate is roughly approximated to the SKG rate, was investigated in \cite{IWCMC_2019}. In this study it was shown that in order to maximize the data rates in the \emph{parallel} approach Alice and Bob should use the strongest subcarriers -- in terms of SNR -- for data transmission and the worst for SKG. Under this simplified formulation, the optimal power allocation for the data transfer has been shown to be a modified waterfilling solution.

Here, we explicitly account for the rate of transmitting reconciliation information and differentiate it from the SKG rate. We confirm whether the policy of using the strongest subcarriers for data transmission and not for reconciliation, is still optimal when the full optimization problem is considered, including the communication cost for reconciliation.

As discussed in Section \ref{subsec:system model}, our physical layer system model assumes Alice and Bob exchange data over a Rayleigh BF-AWGN channel with $N$ orthogonal subcarriers. Without loss of generality the variance of the AWGN in all links is assumed to be unity. During channel probing, constant pilots are sent across all subcarriers \cite{Ye, Belmega} with power $P$. Using the observations (\ref{eq:XAj}), Alice estimates the channel coefficients as 
\begin{align}
{\hat{h}}_j = {h}_j+{\tilde{h}}_j, 
\end{align}
for $j=1, \ldots, N$ where ${\tilde{h}}_j$ denotes an estimation error that can be assumed to be Gaussian, ${\tilde{h}}_j\sim \mathcal{CN}(0, \sigma_e^2)$ \cite{Medard}.
Under this model, the following rate is achievable on the $j$-th subcarrier from Alice to Bob when the transmit power during data transmission is $p_j$ \cite{Medard}:
\begin{align}
    &R_j=\log_2\left(1+\frac{g_jp_j}{\sigma^2_{e} P+1}\right)=\log_2(1+\hat{g}_jp_j),\label{eq:data_rate}
\end{align}
where we use 
\begin{math}
    \hat{g}_i=\frac{{g_i}}{\sigma^2_{i,e} P+1},
\end{math}
\textcolor{black}{to denote the estimated channel gains}. As a result, the channel capacity \begin{math}
    C= \sum_{j=1}^N R_j \label{eq:maxrate}
\end{math} under the short term power constraint:
\begin{equation}
\sum_{j=1}^N p_j \leq NP,\;\; p_j \geq 0,\; \forall j \in\{1, \dots, N\},\label{eq:power}
\end{equation}
is achieved with the well known waterfilling power allocation policy 
\begin{math}
p_j=\left[\frac{1}{\lambda}-\frac{1}{\hat{g}_j}\right]^+, \label{eq:water-filling}
\end{math}
where the water-level $\lambda$ is estimated from the constraint (\ref{eq:power}).
 In the following, the estimated channel gains $\hat{g}_j$ are -- without loss of generality -- assumed ordered in descending order, so that: 
\begin{equation}
\hat{g}_1\geq \hat{g}_2\geq \ldots  \geq \hat{g}_N. \label{eq:ch_gains}
\end{equation}

As mentioned above, the advantage distillation phase of the SKG process consists of the two-way exchange of pilot signals during the coherence time of the channel to obtain $\mathbf{r}_{A,j}, \mathbf{r}_{B,j}, j=1,\ldots,N$. On the other hand, the CSI estimation phase can be used to estimate the reciprocal channel gains in order to optimize data transmission using the waterfilling algorithm. In the former case, the shared parameter is used for generating symmetric keys, in the latter for deriving the optimal power allocation. In the parallel approach the idea is to inter-weave the two procedures and investigate whether a joint encrypted data transfer and key generation scheme as in the AE SKG in Section 4.3 could bear any advantages with respect to the system efficiency. While in the sequential approach the  CSI across all subcarriers will be treated as a source of shared randomness between Alice and Bob, in the parallel approach it plays a dual role.

\subsection{Parallel Approach} \label{subsubsec:parallel}

In the parallel approach, after the channel estimation phase, the legitimate users decide on which subcarrier to send the reconciliation information (e.g., the syndromes as discussed in Section \ref{subsec:system model}) and on which data (\textit{i.e.}, the SKG process here is not performed on all of the subcarriers). The total capacity has now to be distributed between data and reconciliation information bearing subcarriers. As a result, the overall set of orthogonal subcarriers comprises two subsets; a subset $ \mathcal{D}$ that is used for encrypted data transmission with cardinality $|\mathcal{D}| = D$ and a subset $\mathcal{\breve{D}}$ with cardinality $|\mathcal{\breve{D}}|=N-D$ used for reconciliation such that, 
\begin{math}
    \mathcal{D} \cup \breve{\mathcal{D}}=\{1,\ldots,N\}.
\end{math}
Over $\mathcal{D}$ the achievable sum data transfer rate, denoted by $C_D$ is given by
\begin{equation}
C_D=\sum_{j \in \mathcal{D}} \log_2 (1 + \hat{g}_j p_j),  \label{eq:C_D}
\end{equation}
while on the subset $\mathcal{\breve{D}}$, Alice and Bob 
exchange reconciliation information at rate
\begin{equation}
C_R=\sum_{j \in \mathcal{\breve{D}}} \log_2 (1 + \hat{g}_j p_j).
\end{equation}
As stated in Section \ref{subsec:system model} the fading coefficients are assumed to be zero-mean circularly-symmetric complex Gaussian random variables. Using the theory of order statistics, the distribution of the ordered channel gains of the SKG subcarriers, $j \in \breve{\mathcal{D}}$, can be expressed as \cite{Alouini}:
\begin{align}\label{eq:order stats}
f(g_j)\!=\!\frac{N!}{\sigma^2(N-j)!(j-1)!}\!\left(1-e^{-\frac{\hat{g}_j}{\sigma^2}} \right)^{N-j}\!\! \left(e^{-\frac{\hat{g}_j}{\sigma^2}} \right)^{j}\!\!,
\end{align}
where $\sigma^2$ is the variance of the channel gains. As a result of ordering the subcarriers, the variance of each of the subcarriers, is now given by:
\begin{equation}
\sigma_j^2=\sigma^2 \sum_{q=j}^N \frac{1}{q^2}, \quad j \in \{D+1, \dots , N \}.
\end{equation}
Thus, we can now write the SKG rate as (note that the noise variances are here normalized to unity for simplicity) \cite{Ye, Belmega}:
\begin{equation}
    C_{SKG}= \sum_{j\in\breve{\mathcal{D}}}\log_2 \left(1+ \frac{P \sigma_j^2}{2+\frac{1}{P \sigma_j^2}} \right).
\end{equation}

\textcolor{black}{The minimum rate necessary for reconciliation was discussed in Section 4.2. Here, alternatively, we employ a practical design approach in which the rate of the encoder used is explicitly taken into account. Note that  in a rate $\frac{{k}}{{n}}$ block encoder the side information is ${n}-{k}$ bits long, 
i.e., the rate of syndrome to  output key bits after privacy amplification is $\frac{{n}-{k}}{k}$. However, in each key session a 0-RTT look-up identifier of length ${k}$ is also sent. Therefore, we define  the parameter $\kappa=\frac{{n}-{k}}{{k}} + 1=\frac{n}{k}$, i.e., the inverse of the encoder rate, that reflects the ratio of the reconciliation and 0-RTT transmission rate to the SKG rate.  For example, for a rate $\frac{{k}}{{n}}=\frac{1}{2}$ encoder, $\kappa=2$, etc.} Based on this discussion, we capture the minimum requirement for the reconciliation rate through the following expression:
\begin{eqnarray}
C_{R}&\geq&\kappa C_{SKG}.\label{eq:reconciliation} \label{kappa}
\end{eqnarray}

Furthermore, to identify the necessary key rate, we note that depending on the exact choices of the cryptographic suites to be employed, it is possible to reuse the same key for the encryption of multiple blocks of data, e.g., as in the AES GCM, that is being considered for employment in the security protocols for URLLC systems\cite{3gppURLLC}. In practical systems, a single key of length 128 to 256 bits can be used to encrypt up to gigabytes of data. As a result, we will assume that for a particular application it is possible to identify the ratio of key to data bits, which in the following we will denote by $\beta$. Specifically,  we assume that the following security constraint should be met 
    \begin{equation}
{C_{SKG}} \geq \beta C_D, \; \; 0 < \beta \leq 1, \label{eq:security}
\end{equation} 
where, depending on the application, the necessary minimum value of $\beta$ can be identified.  We note in passing that the case $\beta=1$ would correspond to a one-time-pad, \textit{i.e.}, the generated keys could be simply x-ored with the data to achieve perfect secrecy without the need of any cryptographic suites.

Accounting for the reconciliation rate and security constraints in (\ref{eq:reconciliation}) and (\ref{eq:security}) we formulate the following maximization problem:
\begin{eqnarray}
&&\max_{p_j, j\in\mathcal{D}}\sum_{j\in \mathcal{D}} R_{j} \label{eq:optimisation}\\
&&\text{s.t. } (\ref{eq:power}), (\ref{kappa}), (\ref{eq:security}), \nonumber \\ &&\sum_{j\in \mathcal{D}}R_j+\sum_{j\in \mathcal{\breve{D}}}R_j\leq C \label{eq:C}.
\end{eqnarray}
\eqref{eq:security} can be integrated with \eqref{kappa} to the combined constraint
\begin{eqnarray}
\sum_{j\in \mathcal{D}}{R_j}&\leq& \frac{\sum\limits_{j\in\breve{\mathcal{D}}}{R_j}}{\kappa \beta}. \label{eq:combined constraint}
\end{eqnarray}
The optimization problem at hand is a mixed-integer convex optimization problem with unknowns both the sets $\mathcal{D}, \mathcal{\breve{D}}$, as well as the power allocation policy $p_j, j\in \{1,\ldots, N\}$. 
These problems are typically NP hard and addressed with the use of branch and bound algorithms and heuristics.

In this work, we propose a simple heuristic to make the problem more tractable by reducing the number of free variables. In the proposed approach, we assume that the constraint \eqref{eq:C} is satisfied with equality. The only power allocation that allows this is the waterfilling approach that uniquely determines the power allocation $p_j$ and also requires that the constraint (\ref{eq:power}) is also satisfied with equality. Thus, if we follow that approach, we determine the power allocation vector uniquely and can combine the remaining constraints (\ref{eq:C}) and (\ref{eq:combined constraint}) into a single one as:
\begin{equation}
    \sum_{j\in \mathcal{D}}{R_j}\leq \frac{C}{\kappa \beta +1} \label{eq:GAMMA}.
\end{equation}
The new optimization problem can be re-written as\textcolor{black}{ 
\begin{eqnarray}
&&\max_{x_j \in \{0,1\}} \sum_{j = 1}^N{R_j x_j} \label{eq:new optimization} \\
&&\text{s.t. }\sum_{j =1}^N{R_j x_j}  \leq \frac{C}{1+\kappa \beta}. \label{eq:new_constraint}
\end{eqnarray}}
The problem in (\ref{eq:new optimization})-(\ref{eq:new_constraint}) is a subset-sum problem from the family of $0-1$ knapsack problems, that is known to be NP hard~\cite{Knapsack}. However, these type of problems are solvable optimally using dynamic programming techniques  in pseudo-polynomial time \cite{Knapsack, Knapsack_book}. Furthermore, it is known that greedy heuristic approaches are bounded away from the optimal solution by half \cite{Vazirani}. 

We propose a simple greedy heuristic algorithm of \textit{linear complexity,} as follows.\footnote{ \textcolor{black}{Without loss of generality, the algorithm assumes that the channel gains are ordered in decreasing order as in \eqref{eq:ch_gains}, and, consequently, the rates $R_j$ are also ordered in descending order. The ordering is a $\mathcal{O}(N\log N)$ operation and required in common power allocation schemes such as the waterfilling, and, therefore does not come at any additional cost.}} The data subcarriers are selected starting from the best -- in terms of SNR -- until (\ref{eq:new_constraint}) is not satisfied. Once this situation occurs the last subcarrier added to set $\mathcal{D}$ is removed and the next one is added. This continues either to the last index $N$ or until (\ref{eq:new_constraint}) is satisfied with equality. The algorithm is described in \textit{Algorithm 1}.

\restylefloat{algorithm}

\newcommand{\alg}{\texttt{algorithmicx}}
\newcommand{\old}{\texttt{algorithmic}}
\newcommand{\BinSeg}{ Heuristic Greedy Algorithm for (\ref{eq:new optimization})-(\ref{eq:new_constraint})}
\newcommand{\BS}{Heuristic}
\newcommand\ASTART{\bigskip\noindent\begin{minipage}[b]{0.5\linewidth}}
\newcommand\ACONTINUE{\end{minipage}\begin{minipage}[b]{0.5\linewidth}}
\newcommand\AENDSKIP{\end{minipage}\bigskip}
\newcommand\AEND{\end{minipage}}
\alglanguage{pseudocode}
\begin{algorithm}[t]
\caption{\BinSeg}\label{BS}
\begin{algorithmic}[1]
\Procedure{\BS}{start, end, $R_j$}
\State{ $j\leftarrow 1, R_0\leftarrow 0, R_{N+1}\leftarrow 0$}
\While{$j\leq N-1 \text{ and }   \sum_{j =1 }^N{R_j x_j} \leq \frac{C}{1+\kappa \beta}$} 
\State{$\sum_{j =1}^N{R_j x_j} \leftarrow \sum_{j=1}^N{R_{j-1}x_{j-1}}+R_jx_j$}
\If{$\sum_{j=1}^N{R_j x_j}\leq \frac{C}{1+\kappa \beta} $}
\State{$x_j \leftarrow1;j\leftarrow j+1$}
\Else{ do
$x_j\leftarrow 0; j\leftarrow j+1 $}
\EndIf
\EndWhile
\EndProcedure
\end{algorithmic}
\end{algorithm}

The efficiency of the proposed parallel method -- measured as the ratio of the long-term data rate versus the average capacity -- is  evaluated as:
\begin{equation}
  \eta_{\text{parallel}} =\frac{\mathbb{E} \left[\sum\limits_{j\in \mathcal{D}}{R_j}\right]} {\mathbb{E}[C]}. \label{eq:ef-par}
\end{equation}
This efficiency quantifies the expected back-off in terms of data rates when part of the resources (power and frequency) are used to enable the generation of secret keys at the physical layer. In future work, we will compare the efficiency achieved to that of actual approaches currently used in 5G by accounting for the actual delays incurred due to the PKE key agreement operations~\cite{Weinand2018}.

\subsection{Sequential Approach} \label{subsubsec:sequential}
In the sequential approach encrypted data transfer and secret key generation are two separate events; first, the secret keys are generated over the whole set of subcarriers, leading to a sum SKG rate given as  
\begin{equation}
    C_{SKG}=N \log_2 \left(1+ \frac{P \sigma^2}{2+\frac{1}{P \sigma^2}} \right).
\end{equation}
To estimate the efficiency of the scheme, we further need to identify the necessary resources for the exchange of the reconciliation information. We can obtain an estimate of the number of transmission frames that will be required for the transmission of the syndromes, as the expected value of the reconciliation rate (\textit{i.e.}, it's long-term value) $\mathbb{E}[C_R]$. The average number of frames needed for reconciliation is then computed as:
\begin{equation}
    M=\ceil[\Bigg]{\frac{\kappa C_{SKG}}{\mathbb{E}[C_R]} },
\end{equation}
where $\ceil x $ 
denotes the smallest integer that is larger than $x$.

 The average number of the frames that can be sent while respecting the secrecy constraint is:
\begin{equation}
    L= \floor[\Bigg]{ \frac{C_{SKG}}{\beta\mathds{E}[C]}} ,
\end{equation}
where $\floor x $ 
denotes the largest interger that is smaller than $x$. 
The efficiency of the sequential method is then calculated as:
\begin{equation}
    \eta_{\text{sequential}}=\frac{L}{L+M}. \label{eq:ef-seq}
\end{equation}

\section{Effective Data Rate Taking into Account Statistical Delay QoS Requirements}
\label{sec:eff_capacity}
\textcolor{black}{ In the previous section, we investigated the optimal power and subcarrier allocations strategy of Alice and Bob in order to maximize their long-term average data rate and proposed a 
greedy heuristic algorithm of linear complexity. 
Here, we extend our work from Section \ref{sec:hybrid} by taking into account delay requirements. In detail, we investigate the optimal resource allocation for Alice and Bob, when their communication has to satisfy specific delay constraints. To this end, we use the theory of \textit{effective capacity} \cite{Negi} which gives a limit for the maximum arrival rate under delay-bounds with a specified violation probability.}

We study the \textit{effective data rate} for the proposed pipelined SKG and encrypted data transfer scheme; the effective rate is a data-link layer metric that captures the impact of  statistical delay QoS constraints on the transmission rates. As background, we refer to \cite{Queue} which showed that the probability of a steady-state queue length process $Q(t)$ exceeding a certain queue-overflow threshold $x$ converges to a random variable $Q(\infty)$ as:
\begin{equation}
     \lim_{x \rightarrow \infty} \frac{\ln(\text{Pr}[Q (\infty) > x])}{x}= -\theta, \label{eq:queue_theta}
\end{equation}
where $\theta$ indicates the asymptotic exponential decay-rate of the overflow probability. For a large threshold $x$, \eqref{eq:queue_theta} can be represented as $Pr[Q (\infty) > x] \approx e^{-\theta x}$. Furthermore, the delay-outage probability can be approximated by \cite{Negi} :
\begin{equation}
    \text{Pr}_{\text{delay}}^{\text{out}}\!\! =\!\! \text{Pr}[\text{Delay}\! > \!\! D_{\text{max}}] \! \approx \! \text{Pr}[Q(\infty)\!\!>\! 0]e^{-\theta \zeta D_{\text{max}}}, \label{eq:delay_outage}
 \end{equation}
where $D_{\text{max}}$ is the maximum tolerable delay, \(\text{Pr}[Q(\infty)> 0]\) is the probability of a non-empty buffer, which can be estimated from the ratio of the constant arrival rate to the averaged service rate,  $\zeta$ is the upper bound for the constant arrival rate when the statistical delay metrics are satisfied. 

Using the delay exponent ($\theta$) and the probability of non-empty buffer, the effective capacity, that denotes the maximum arrival rate, can be formulated as \cite{Negi}: 
\begin{equation}
    E_C(\theta) = - \lim_{t \rightarrow \infty} \frac{1}{\theta}\ln \mathbb{E} [e^{-\theta S[t]}] (\text{bits/s}),
 \end{equation}
where \(S[t] = \sum_{i=1}^{t} s[i]\) denotes the time-accumulated service process, and {$s[i], i = 1, 2, . . .$} denotes the discrete-time stationary and ergodic stochastic service process. Therefore, the delay exponent $\theta$ indicates how strict the delay requirements are, \textit{i.e.}, $\theta \rightarrow 0$ corresponds to looser delay requirements, while $\theta \rightarrow \infty$ implies exceptionally stringent delay constraints. Assuming a Rayleigh block fading system, with frame duration $T_f$ and total bandwidth $B$, we have $s[i]=T_f B\tilde{R}_i$, with $\tilde{R}_i$ representing the instantaneous service rate achieved during the duration of the $i^{th}$ frame. In the context of the investigated data and reconciliation information transfer, $\tilde{R}_i$, is given by:
\begin{align}\label{eq:rate}
    \tilde{R}_i&=\frac{1}{F}\sum_{i\in\mathcal{D}}{\log_2 (1+p_i \hat{g}_i)},
\end{align}
where $F$ is the equivalent frame duration, \textit{i.e.}, the total number of subcarriers used for data transmission, so that for the parallel approach we have $F=|D|$ while for the sequential approach $F=N(L+M)L^{-1}$. 


  \begin{figure*}[t]
  \setcounter{figure}{3}
    \begin{multicols}{2}
\centering
  \includegraphics[clip, trim=3.5cm 9.4cm 4cm 10cm, width=0.48\textwidth]{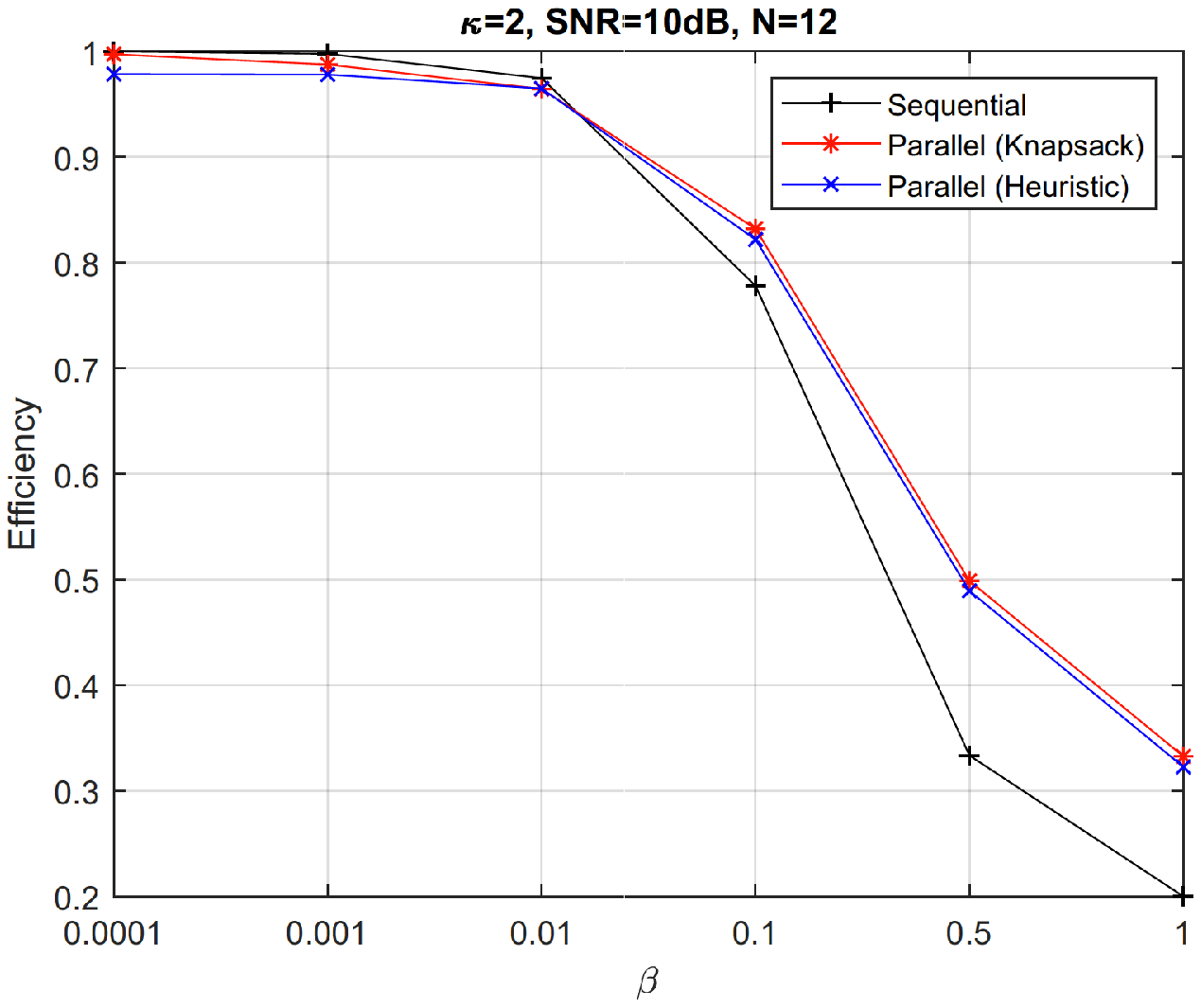} 
  \caption{\textbf{a)} Efficiency comparison  for $N=12$, SNR=$10$ dB and $\kappa=2$.} \label{fig:1}
  \setcounter{figure}{3}
  \includegraphics[clip, trim=3.5cm 9.4cm 4cm 10cm, width=0.48\textwidth]{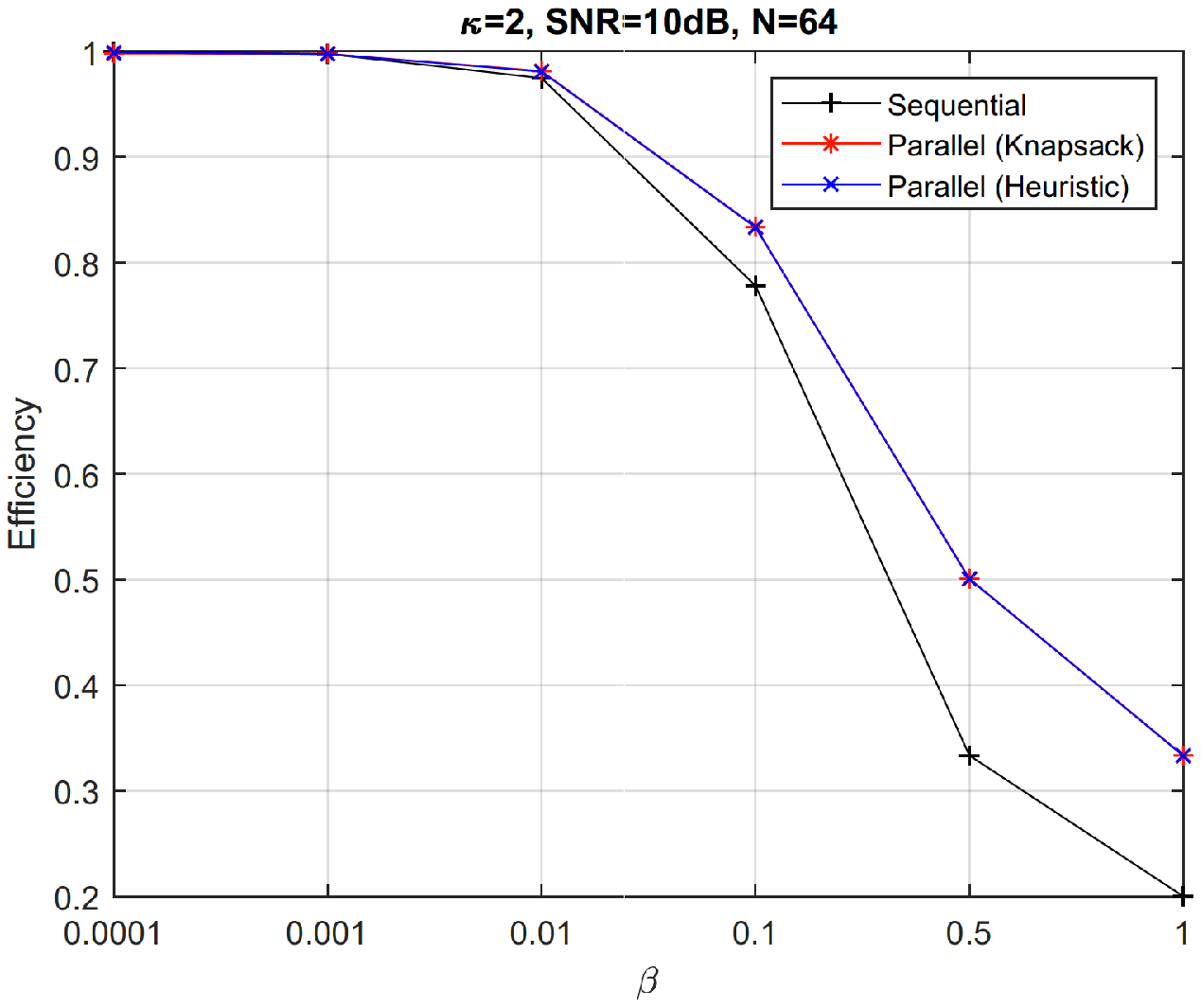} 
  \caption{\textbf{b)} Efficiency comparison  for $N=64$,  SNR=$10$ dB and $\kappa=2$.} \label{fig:2}
  \end{multicols}
   \end{figure*}

    \begin{figure}[t]
  \includegraphics[width=0.51\textwidth]{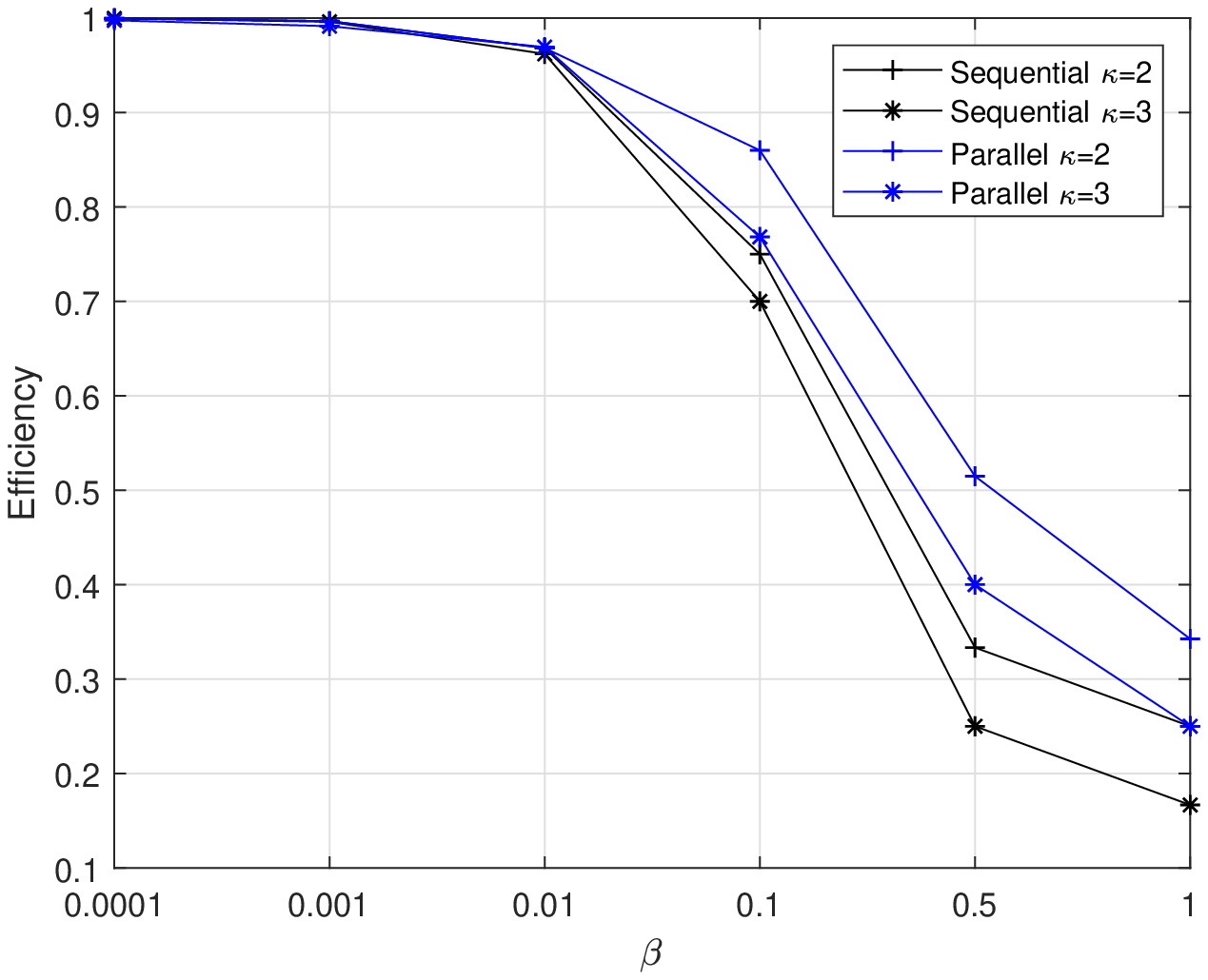}
  \caption{Efficiency vs $\kappa$, for $N=24$, SNR=$10$ dB.} \label{fig:eff_kappa}
  \end{figure}

Under this formulation and assuming that  G\"{a}rtner-Ellis theorem \cite{Gartner, Ellis} is satisfied, the \emph{effective data rate}\footnote{Since part of the transmission rate is used for reconciliation information, and part for data transmission the terms ``\emph{effective syndrome rate}'' and ``\emph{effective data rate}'' are introduced instead of the term ``effective capacity'', for rigour. {We note that we assume the information data and reconciliation information are accumulated in separate independent buffers within the transmitter. } } $E_C(\theta)$ is given as:
\begin{equation}
    E_{C,\mathcal{D}}(\theta)=- \frac{1}{\theta T_f B} \ln \left(\mathbb{E} \left[e^{-\theta T_f B  \tilde{R}_i}\right]\right). \label{eq:eff_cap}
\end{equation}
We set $\alpha=\frac{\theta T_f B}{\ln(2)}$.
By inserting \eqref{eq:rate} into \eqref{eq:eff_cap} we get:
\begin{align}
    E_{C,\mathcal{D}}(\theta)&\!= \! \!-\frac{1}{\ln(2)\alpha}\!\ln\! \left(\mathbb{E} \! \left[e^{-\ln(2)\alpha  F^{-1} \!\sum_{i \in \mathcal{D}}\!\log_2 \left(1+ p_i \hat{g}_i \right)}\right]\!\right)\!, \nonumber\\
E_{C,\mathcal{D}}(\theta)&\!=-\frac{1}{\alpha}\log_2 \left(\mathbb{E} \left[ \prod_{i \in \mathcal{D}}\left(1+ p_i \hat{g}_i \right)^{-\alpha F^{-1} }\right]\right). \label{eq:Ec_before}
\end{align}
Assuming i.i.d. channel gains, by using the distributive property of the mathematical expectation, \eqref{eq:Ec_before} becomes \cite{Taufik}:\textcolor{black}{
\begin{align}
    E_{C,\mathcal{D}}(\theta)\!=-\frac{1}{\alpha}\log_2 \left( \prod_{i \in \mathcal{D}} \mathbb{E} \left[ \left(1+ p_i \hat{g}_i \right)^{-\alpha F^{-1} }\right]\right).
\end{align}
We further manipulate by using the log-product rule to obtain:
\begin{align}
E_{C\mathcal,{D}}(\theta)=-\frac{1}{\alpha} \sum_{i \in \mathcal{D}}\log_2 \left(\mathbb{E} \left[\left(1+ p_i \hat{g}_i \right)^{-\alpha F^{-1} }\right]\right).\label{eq:eff_rate}
\end{align}}
Similarly, the \emph{effective syndrome rate} can be written as:
\begin{equation}
    E_{C\breve{\mathcal,{D}}}(\theta)=-\frac{1}{\alpha} \sum_{i \in \breve{\mathcal{D}}}\log_2 \left(\mathbb{E} \left[\left(1+ p_i \hat{g}_i \right)^{-\alpha \breve{F}^{-1} }\right]\right),\label{eq:eff_rec_rate}
\end{equation}
where the size of  $\breve{F}$ here is $|N-D|$.

Using that, we now reformulate the maximization problem given in \eqref{eq:optimisation} by adding a delay constraint. The reformulated problem can be expressed as follows: \begin{eqnarray}
&&\max_{p_j, j\in\mathcal{D}}  E_{C,\mathcal{D}}(\theta), \label{eq:opt_problem_EC}\\
&&\text{s.t. } (\ref{eq:power}), (\ref{eq:combined constraint}), \nonumber \\ &&E_{C,\mathcal{D}}(\theta)+E_{C,\breve{\mathcal{D}}}(\theta)\leq E_{C}^{\text{opt}}(\theta), \label{eq:E_C_constraint}
\end{eqnarray}
where $E_{C}^{\text{opt}}(\theta)$ represents the maximum achievable effective capacity for both key and data transmission for a given value of $\theta$ over $N$ subcarriers:
\textcolor{black}{\begin{align}
E_{C}^{\text{opt}}(\theta)\! =\!\!\!\!\!\!\!\!\!\max_{p_i, i=1,2, \dots N} \!\!\left\{\!-\frac{1}{\alpha}\log_2 \left(\!\mathbb{E}\! \left[ \prod_{i=1}^N\! \left(1+ p_i \hat{g}_i \right)^{-\alpha N^{-1} }\!\right]\!\right)\!\right\}. \label{eq:EC_optimal}
\end{align}}

In the proposed approach, we assume that the constraint \eqref{eq:E_C_constraint} is satisfied with equality. Given that, \textcolor{black}{the optimization problem in \eqref{eq:opt_problem_EC} can be evaluated as two sub-optimization problems: i) finding the optimal long term power allocation from \eqref{eq:power} and \eqref{eq:EC_optimal}; ii) finding the optimal subcarrier allocation that satisfies \eqref{eq:combined constraint}. We solve the first problem that gives the optimal power allocation using convex optimization tools. Next, as in Section \ref{sec:hybrid} we use two methods to solve subcarrier allocation problem, i.e., by formulating a subset-sum 0 -- 1 knapsack optimization problem or through a variation of \textit{Algorithm 1}. The efficiency of both methods is compared numerically to the sequential method in Section \ref{sec:results}.} 


\textcolor{black}{Now, following the same steps as in \eqref{eq:Ec_before}-\eqref{eq:eff_rate} and using  the fact that maximizing $E_C(\theta)$ is  equivalent  to  minimizing $-E_C(\theta)$ (this is due to $\log(\cdot)$ being a monotonically increasing concave function for any $\theta>0$) we formulate the following minimization problem:}
\begin{align}
   &\min_{p_i, i=1,2, \dots N} \sum_{i=1}^N \left(\mathbb{E} \left[\left(1+ p_i \hat{g}_i \right)^{-\alpha N^{-1} }\right]\right), \label{eq:minimise} \\
   &\text{ s.t. $\eqref{eq:power}$}. \nonumber 
\end{align} 
where $F=N$ in this case as the full set of subcarriers is concerned. We form the Lagrangian function $\mathcal{L}$ as:
\begin{align}
    \mathcal{L} =  \left(\mathbb{E} \left[\left(1+ p_i \hat{g}_i \right)^{-\alpha N^{-1} }\right]\right) + \lambda \left(\sum_{i=1}^N p_i - NP\right). \label{eq:Langrangian}
\end{align}
By differentiating \eqref{eq:Langrangian} w.r.t. $p_i$ and setting the derivative equal to zero \cite{Convex_optimization} we get:
\begin{align}
\frac{\partial \mathcal{L}}{\partial p_i} = \lambda- \frac{\alpha \hat{g}_i}{N} \left(\hat{g}_i p_i+1\right)^{ - \frac{\alpha}{N}- 1} = 0. \label{eq:zero}
\end{align}
Solving \eqref{eq:zero} gives the optimal power allocation policy:
\begin{equation}
    p_i^*=\frac{1}{g_0^{\frac{N}{\alpha+N}}\hat{g}_i^{\frac{\alpha}{\alpha+N}}}- \frac{1}{\hat{g}_i}, \label{eq:p_star}
\end{equation}
where $g_0=\frac{N \lambda}{\alpha}$ is the cutoff value which can be found from the power constraint. By inserting $p_i^*$ in $E_C(\theta)$ we obtain the expression for $E_C^{\text{opt}}(\theta)$:
\begin{equation}
    E_{C}^{\text{opt}}(\theta)=-\frac{1}{\alpha} \sum_{i=1}^N\log_2 \left(\mathbb{E} \left[\left(\frac{\hat{g}_i}{g_0} \right)^{-\frac{\alpha}{\alpha+N}}\right]\right)
\end{equation}
When $\theta \rightarrow 0$ the optimal power allocation is equivalent to water-filling and when $\theta \rightarrow \infty$ the optimal power allocation transforms to total channel inversion.

Now, fixing the power allocation as in \eqref{eq:p_star} we can easily find the optimal subcarrier allocation that satisfies \eqref{eq:combined constraint}. As in Section \ref{sec:hybrid} to do that we first formulate a subset-sum 0 -- 1 knapsack optimization problem that we solve using the standard dynamic programming approach. Furthermore we evaluate the performance of the heuristic algorithm presented in \textit{Algorithm 1}.

  \begin{figure*}[t]
  \begin{multicols}{2}
 \centering
 \includegraphics[clip, trim=3.5cm 9.4cm 4cm 10cm, width=0.45\textwidth]{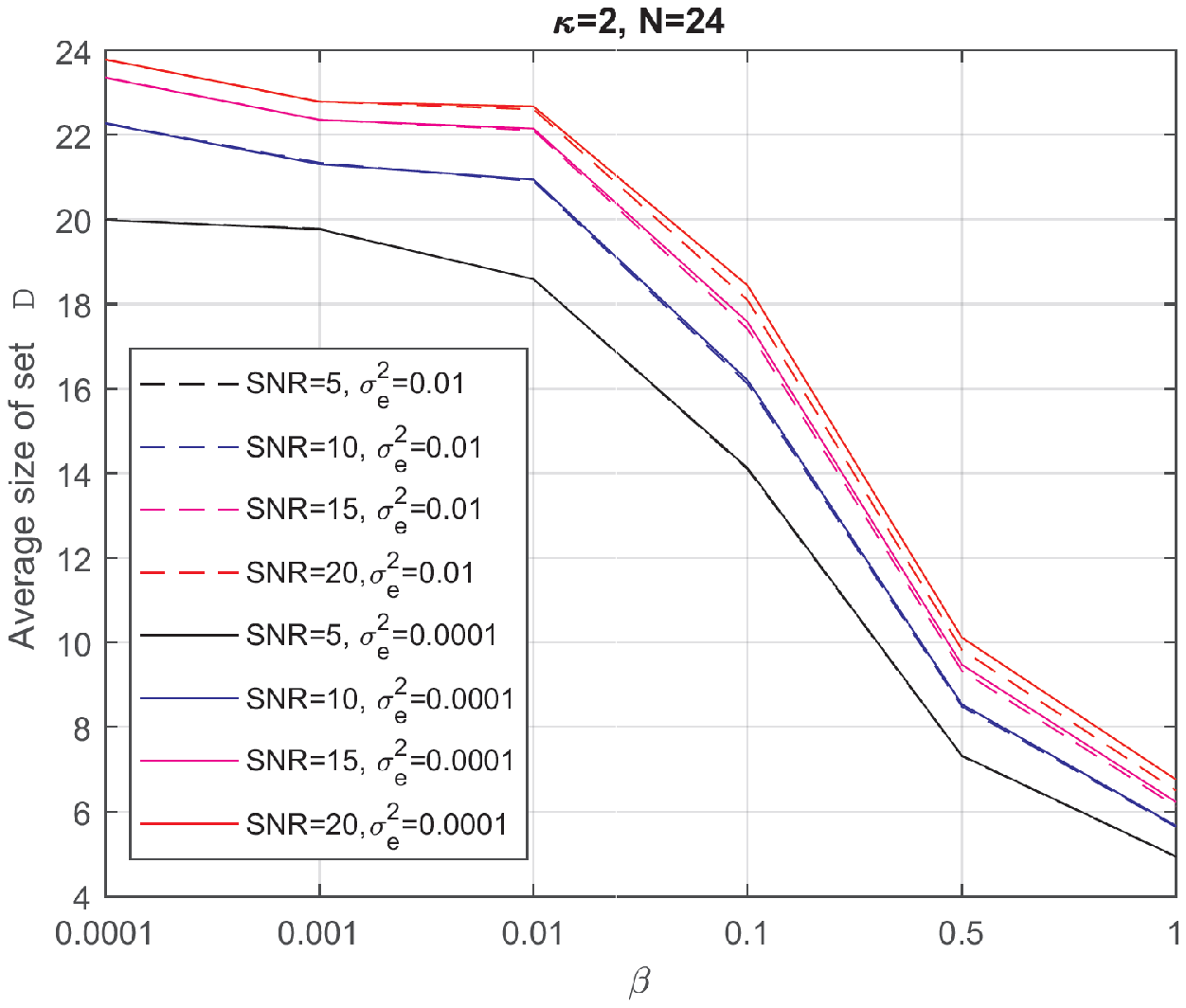} 
 \caption{\textbf{a)} Size of set $\mathcal{D}$ for different SNR levels and $\sigma_e^2$ when $N=24.$}  \label{fig:size_D_SNR}
   \setcounter{figure}{5}
 \includegraphics[clip, trim=0cm 0cm 0cm 0.6cm, width=0.49\textwidth]{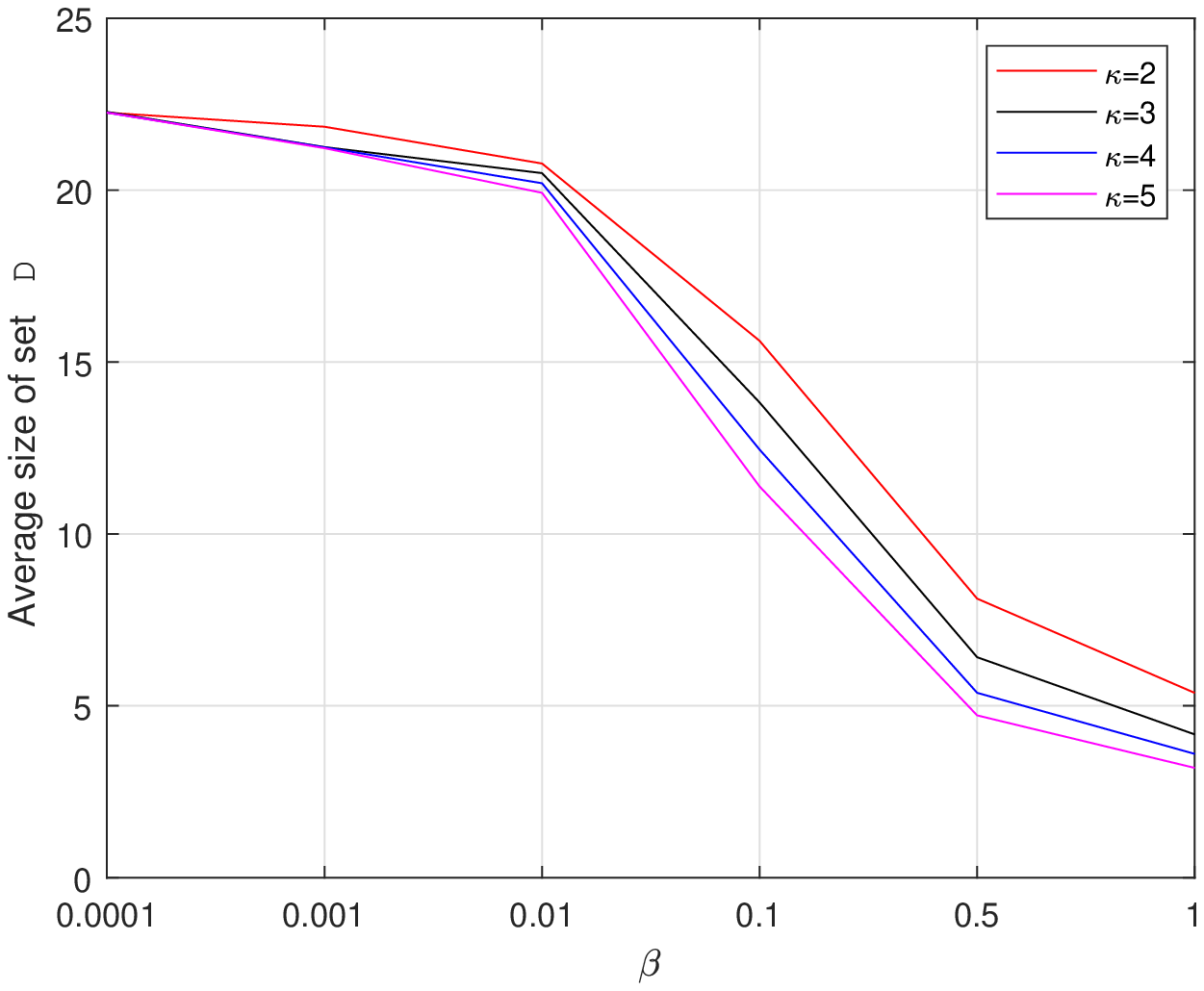} 
\caption{\textbf{b)} Size of set $\mathcal{D}$ for different values of $\kappa$ when $N=24.$} \label{fig:size_D_kappa}  
 \label{fig:size_D}
 \end{multicols}
   \end{figure*}


\section{Results and Discussion}
\label{sec:results}

In this Section we provide numerical evaluations of the efficiency that can be achieved with the presented methods (\textit{i.e.}, sequential and parallel) for different values of the main parameters. With respect to the parallel approach, we provide numerical results of the optimal dynamic programming solution of the subset-sum $0-1$ knapsack problem, as well as of the greedy heuristic approach presented in \textit{Algorithm 1}. \textcolor{black}{For the case of the long term average data rate $C_D$ \eqref{eq:C_D}, we compare the two methods through their efficiencies, \textit{i.e.} $\eta_{\text{sequential}}$ and $\eta_{\text{parallel}}$ given in \eqref{eq:ef-seq} and \eqref{eq:ef-par}, respectively. Next, to compare the two methods in the case of \emph{effective data rate} we evaluate $E_{C\mathcal,{D}}(\theta)$ given in \eqref{eq:eff_rate}.} For better illustration of each case they are separated into different subsections.

\subsection{Numerical results for the case long term average $C_D$}
 
Figures \ref{fig:1}a and \ref{fig:2}b show the efficiency of the methods for $N=12$, and $N=64$, respectively, while $\kappa=2$ and $P=10$. We note  that  the proposed heuristic algorithm has a near-optimal performance (almost indistinguishable from the red curves  achieved with dynamic programming). Due to this fact (which was tested across all scenarios that follow) only the heuristic approach is shown in  subsequent figures for clarity in the graphs. 

We see that when there are a small number of subcarriers ($N$=12, typical for NB-IoT) and small $\beta$ the efficiency of both the parallel \textcolor{black}{$\eta_{\text{parallel}}$} and  the sequential \textcolor{black}{$\eta_{\text{sequential}}$} approaches are very close to unity, a trend that holds for increasing $N$.  
With increasing $\beta$, due to the fact that more frames are needed for reconciliation in the sequential approach (\textit{i.e.}, $M$ increases), regardless of the total number of subcarriers, the parallel method   proves more efficient than the sequential. While the efficiency of the sequential and parallel methods coincide almost until around $\beta=0.01$ for $N=12$, for $N=64$ the crossing point of the curves moves to the left and the efficiency of the two methods coincide until around $\beta=0.001$. This trend was found to be consistent across many values of $N$, only two of which are shown here for compactness of presentation. 

Next, in Fig. \ref{fig:eff_kappa} the efficiency of the parallel \textcolor{black}{$\eta_{\text{parallel}}$} and the sequential \textcolor{black}{$\eta_{\text{sequential}}$} methods are shown for two different values of $\kappa \in \{2, 3\}$ for SNR $=10$ dB and $N=24$. It is straightforward to see that they both follow similar trends and when $\kappa$ increases the efficiency decreases. On the other hand, regardless of the value of $\kappa$ they both perform identically until around $\beta=0.001$.

Finally, in Fig. 4, focusing on the parallel method,  the average size of set $\mathcal{D}$ is shown for different values of $\sigma_e^2$ and SNR levels (Fig. \ref{fig:size_D_SNR}a) and $\kappa$ (Fig. \ref{fig:size_D_kappa}b), for $N=24$. As expected, in Fig. \ref{fig:size_D_SNR}a we see when the SNR increases the size of the set increases, too. This is due to the fact that more power is used on any single subcarrier and consequently a higher reconcilliation rate can be sustained. Regarding the estimation error $\sigma^2_e$ of the CSI, it only slightly affects the performance at high SNR levels. Hence more subcarriers have to be used for reconciliation, and fewer for data. 
The SNR level in Fig. \ref{fig:size_D_kappa}b is set to $10$  dB. The figure shows that when increasing $\kappa$ the size of set $\mathcal{D}$ decreases. This result can be easily predicted from inequality \eqref{eq:reconciliation}, meaning, when $\kappa$ increases more reconciliation data has to be sent, hence fewer subcarriers can be used for data. In both Fig. \ref{fig:size_D_SNR}a and Fig. \ref{fig:size_D_kappa}b when $\beta$ increases the size of set $\mathcal{D}$ decreases; this effect is a consequence of constraint \eqref{eq:new_constraint} as the data rate is decreasing with $\beta$.

\subsection{Numerical results for the case of \emph{effective data rate}}

  \begin{figure*}[t]
  \begin{multicols}{2}
  \setcounter{figure}{6}
  \includegraphics[width=0.48\textwidth]{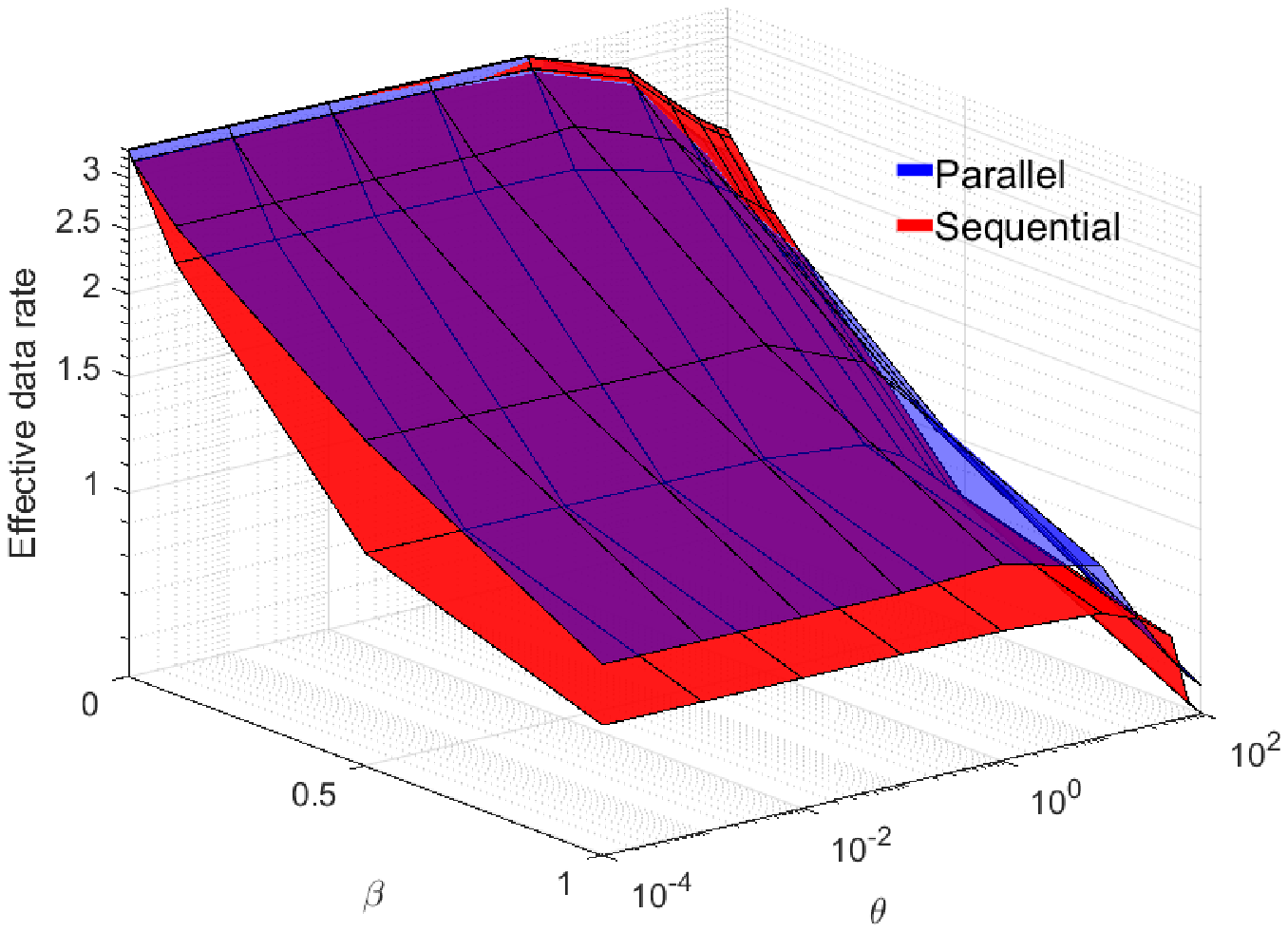} 
\caption{\textbf{a)} Effective data rate achieved by the parallel heuristic approach and the sequential approach when $N=12$, SNR$=10$ dB and $\kappa=2$.}  
\label{fig:surfn12}
\setcounter{figure}{6}
  \includegraphics[width=0.48\textwidth]{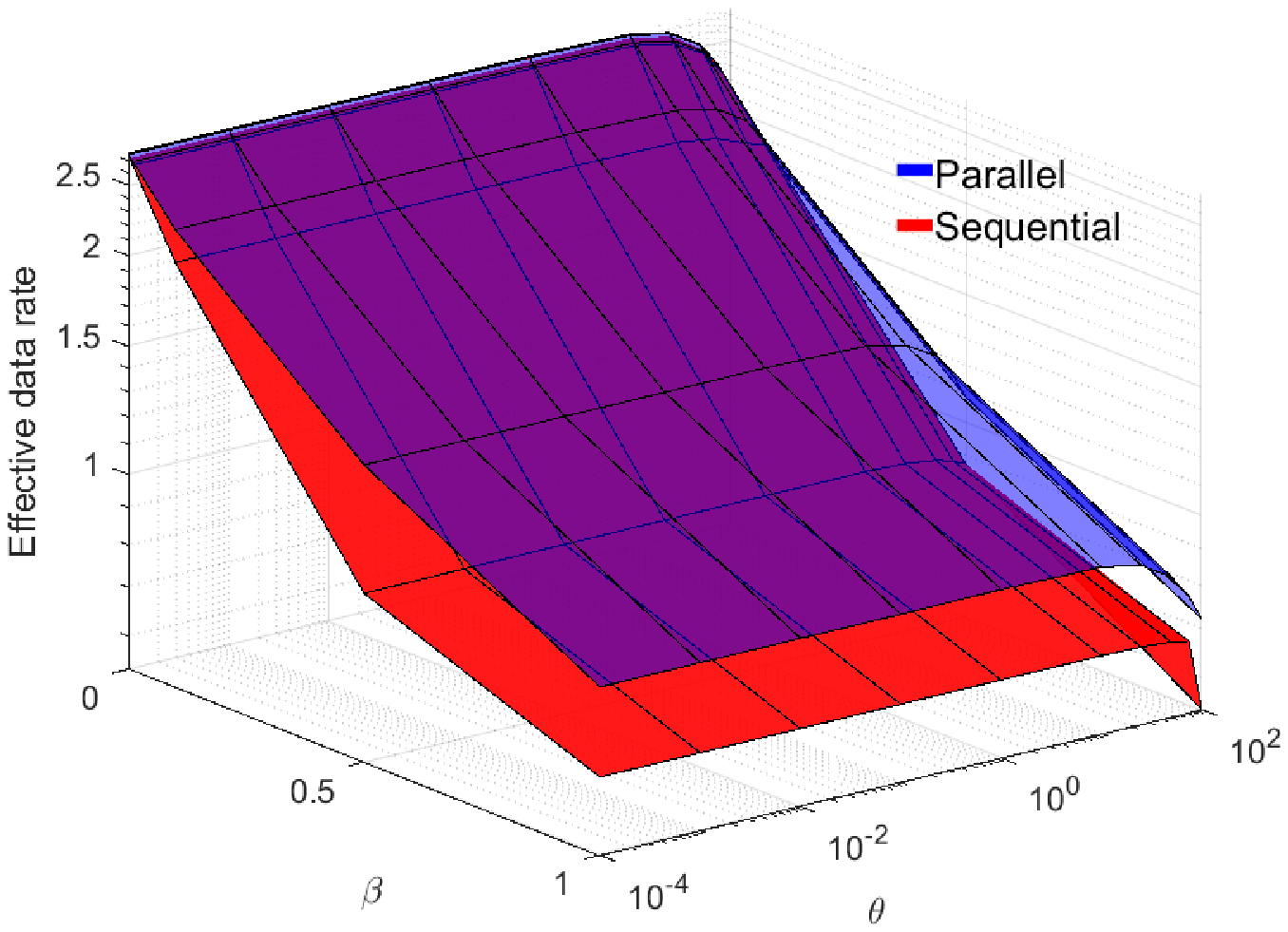} 
\caption{\textbf{b)} Effective data rate achieved by the parallel heuristic approach and the sequential approach when $N=64$, SNR$=10$ dB and $\kappa=2$.}
\label{fig:surfn64} 
  \end{multicols}
  \end{figure*}

\setcounter{figure}{6}
  \begin{figure*}[t]
  \begin{multicols}{2}
  \includegraphics[width=0.48\textwidth]{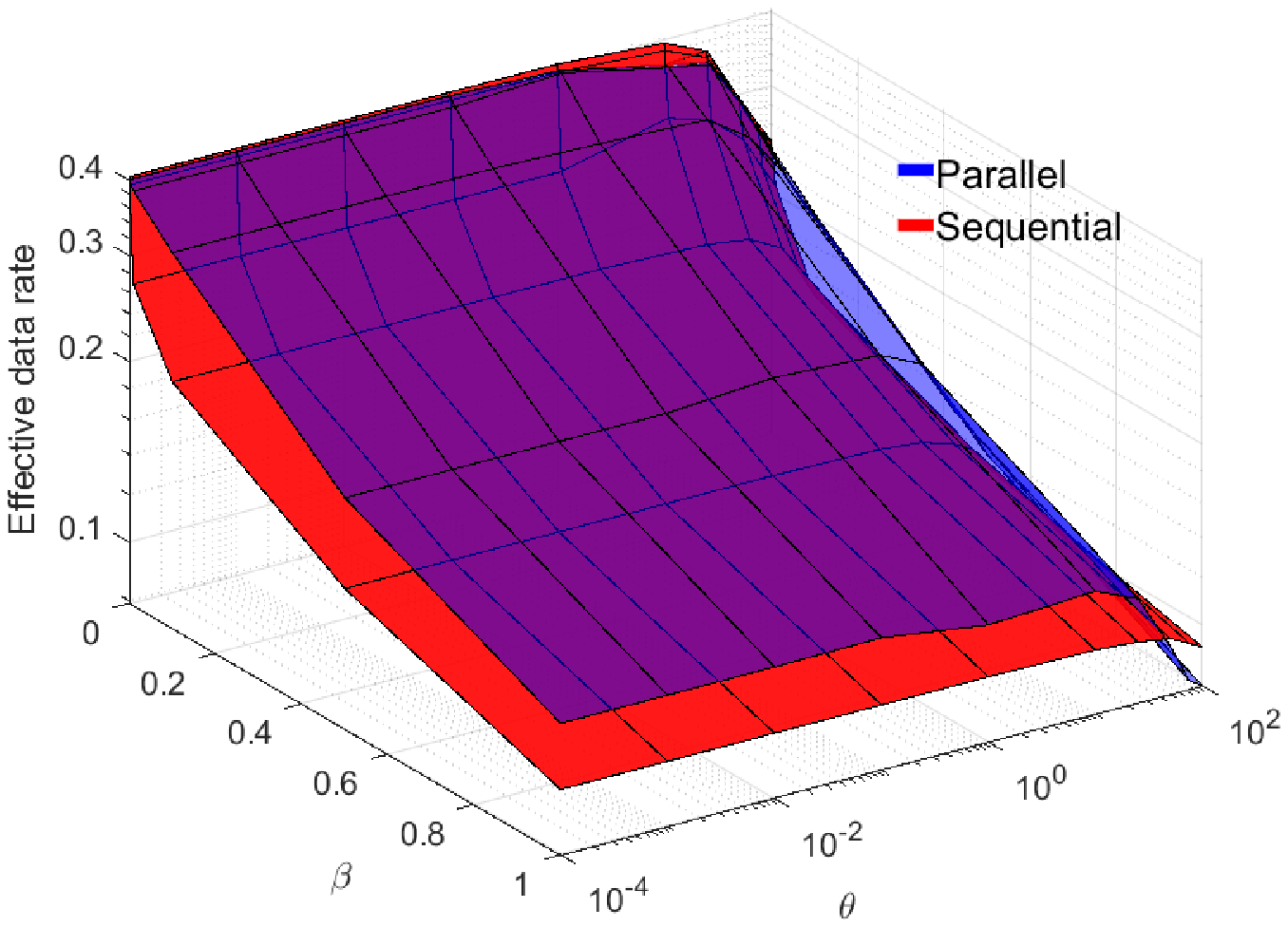} 
\caption{\textbf{c)} Effective data rate achieved by the parallel heuristic approach and the sequential approach when $N=12$, SNR$=0.2$ dB and $\kappa=2$.}  
\label{fig:surfn12p02}
\setcounter{figure}{6}
  \includegraphics[width=0.48\textwidth]{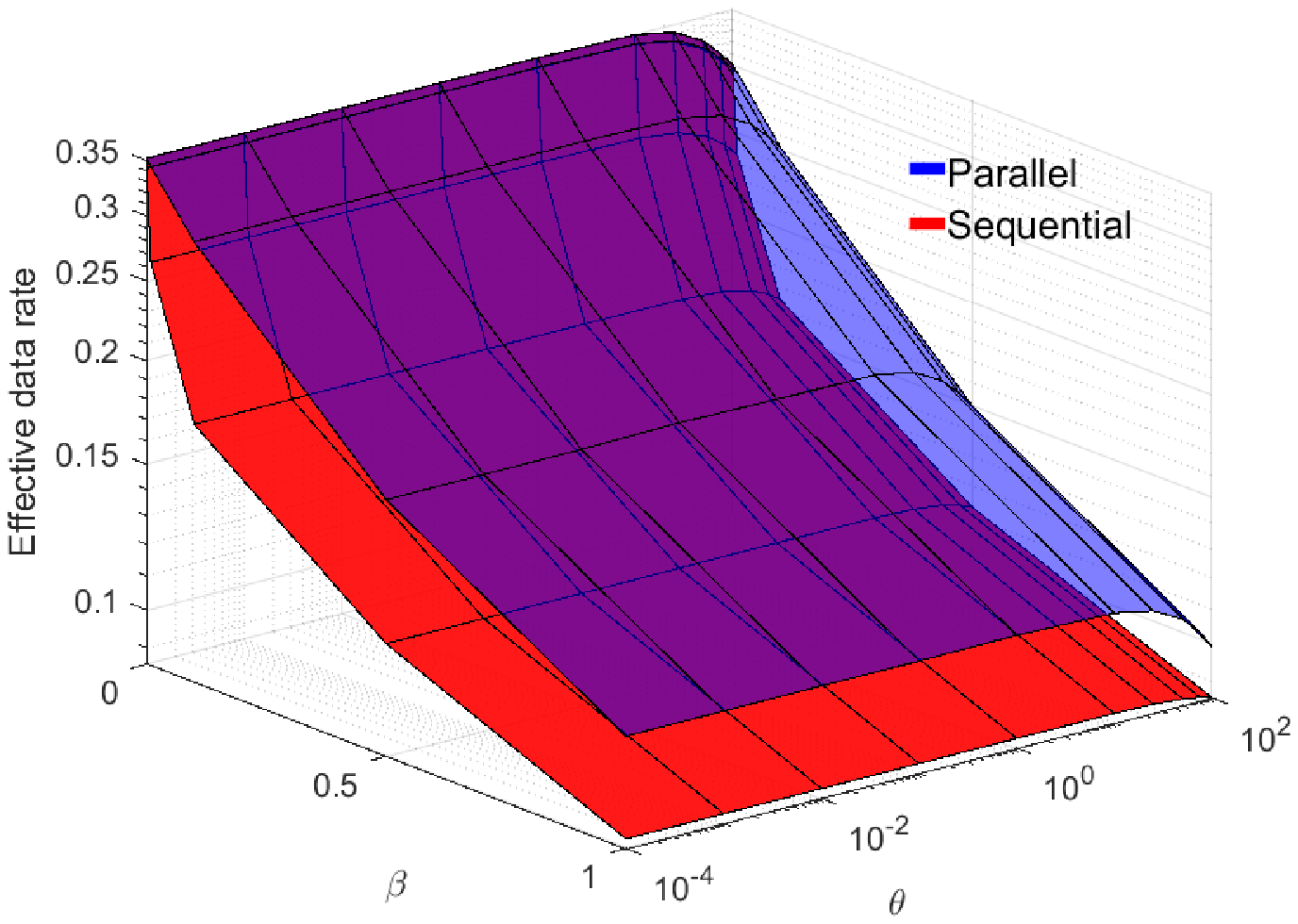} 
\caption{\textbf{d)} Effective data rate achieved by the parallel heuristic approach and the sequential approach when $N=64$, SNR$=0.2$ dB and $\kappa=2$.}
\label{fig:surfn64p02} 
  \end{multicols}
  \end{figure*}
  
\textcolor{black}{ Inspired by the good performance of \emph{Algorithm 1}, in the case where long-term average rate is the metric of interest, here, we continue our investigation with a variation of \emph{Algorithm 1}, with the following differences: at lines 3 and 5 instead of \eqref{eq:GAMMA} we use the constraint \eqref{eq:combined constraint}, the power allocation is fixed as in \eqref{eq:p_star}. The performance of our system is again compared with a sequential method and the metric of interest here is the \emph{effective data rate}. The comparison is performed by taking into account the following parameters: signal to noise ration (SNR); number of subcarriers $N$; ratio of the reconciliation and $0-$RTT transmission rate to the SKG rate $\kappa$; delay exponent $\theta$; and, the ratio of key bits to data bits $\beta$.  }

In Fig. \ref{fig:surfn12} we give a three-dimensional plot showing the dependence of the achievable \emph{effective data rate} $E_{C\mathcal,{D}}(\theta)$ on $\beta$ and $\theta$. Figures \ref{fig:surfn12}a and \ref{fig:surfn64}b compare the parallel heuristic approach and the sequential approach for high SNR levels, whereas Fig. \ref{fig:surfn12p02}c and \ref{fig:surfn64p02}d compare their performance for low SNR level. In  Fig. \ref{fig:surfn12}a and \ref{fig:surfn12p02}c we have $N=12$ while in Fig. \ref{fig:surfn64}b and \ref{fig:surfn64p02}d the total number of subcarriers is $N=64$. All graphs compare the performance of the heuristic parallel approach and the sequential approach for $\kappa=2$.

\textcolor{black}{As discussed in Section \ref{sec:eff_capacity}, when the delay exponent $\theta$ increases,  the optimal power allocation transforms from waterfilling to total channel inversion. Consequently, the rate achieved on all subcarriers converges to the same value, hence when we a have small number of subcarriers (such as $N= 12$) and small values of $\beta$ then using a single subcarrier for reconciliation data will use more capacity than needed and most of the rate on this subcarrier is wasted. Devoting a whole subcarrier for sending the reconciliation data for the case of $N=12$ and $\beta=0.0001$ is almost equivalent of losing $1/12$ of the achievable rate.}  

This can be seen for in Fig. \ref{fig:surfn12}a and \ref{fig:surfn12p02}c. When the SNR is high (See Fig. \ref{fig:surfn12}a), as discussed, this effect is mostly noticeable for large values of $\theta$ and small values of $\beta$\footnote{\emph{i.e} that the ratio of reconciliation information to data is small as seen from Eq. \eqref{eq:combined constraint})}, whereas for small values of $\beta$ and $\theta$ both algorithms perform nearly identically. A similar trend can be seen at the low SNR regime in Fig. \ref{fig:surfn12p02}c. However, at a low SNR the sequential approach has a lower effective data rate. This happens because at high SNR levels each reconciliation frame will contain more information and hence more data frames will follow. Therefore, at the low SNR regime, the reconciliation information received will decrease, hence less data can be sent afterwards. This does not affect the parallel approach.
However, in both scenarios high SNR Fig. \ref{fig:surfn12}a and low SNR Fig. \ref{fig:surfn12p02}c, when $\beta$ increases regardless of the value of $\theta$ the parallel approach always achieves higher \emph{effective data rate} $E_{C\mathcal,{D}}(\theta)$.

In the next case, when the total number of subcarriers is $N=64$, illustrated in Fig. \ref{fig:surfn64}b and \ref{fig:surfn64p02}d, we see that the penalty of devoting a high part of the achievable effective capacity $E_{C}^{\text{opt}}(\theta)$ to reconciliation disappears and the heuristic parallel approach always achieves higher or identical \emph{effective data rate} $E_{C\mathcal,{D}}(\theta)$ compared to the sequential approach. This trend repeats for high and low SNR levels as given in Fig. \ref{fig:surfn64}b and \ref{fig:surfn64p02}d, respectively.


  \begin{figure*}[t]
  \begin{multicols}{2}
  \setcounter{figure}{7}
  \includegraphics[width=0.48\textwidth]{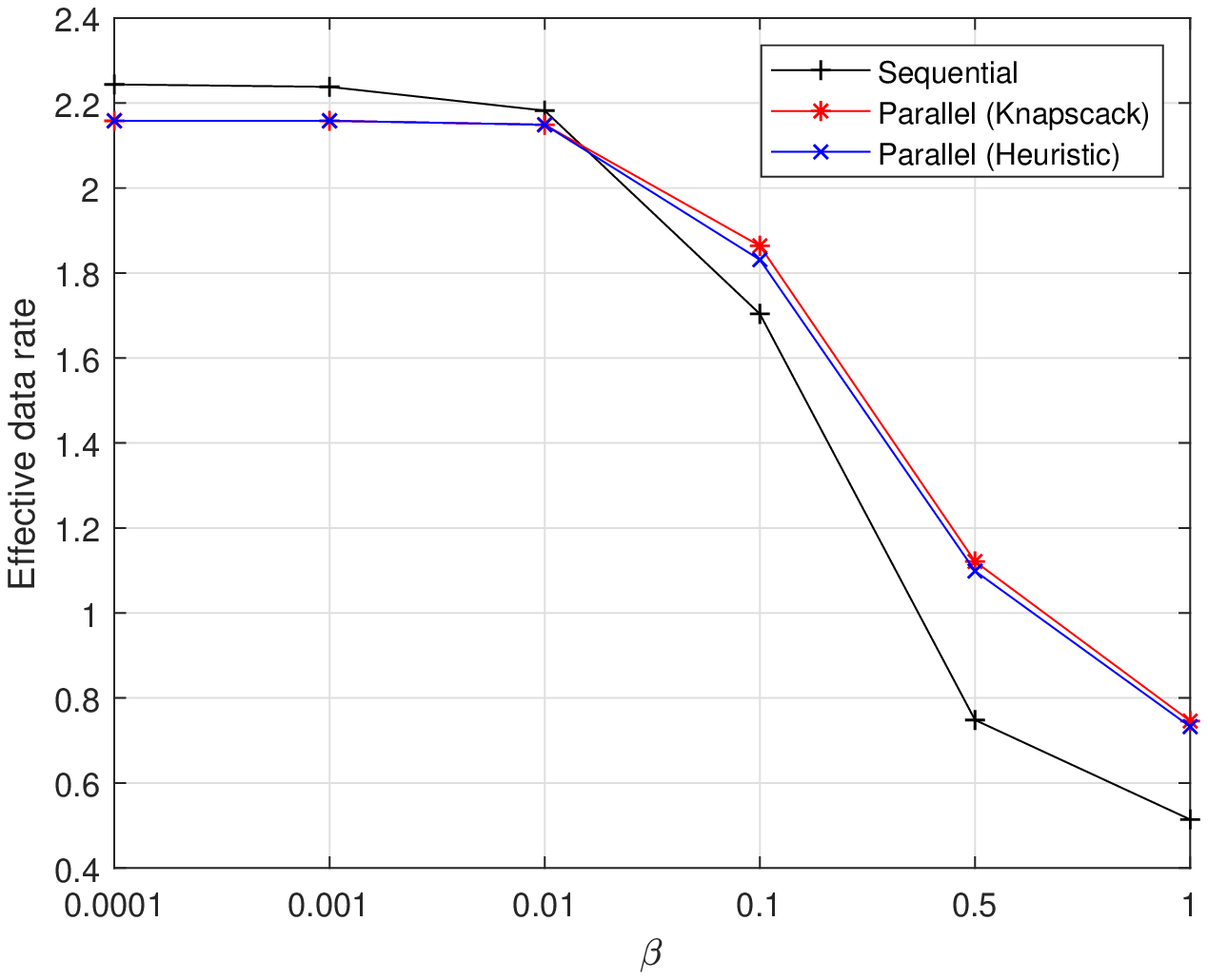}
\caption{\textbf{a)} Effective data rate achieved by parallel and sequential approaches when $N=12$, SNR$=5$dB, $\theta=0.0001, \kappa=2$.} 
\label{fig:n12_smalltheta}  
 \setcounter{figure}{7}
    \includegraphics[width=0.48\textwidth]{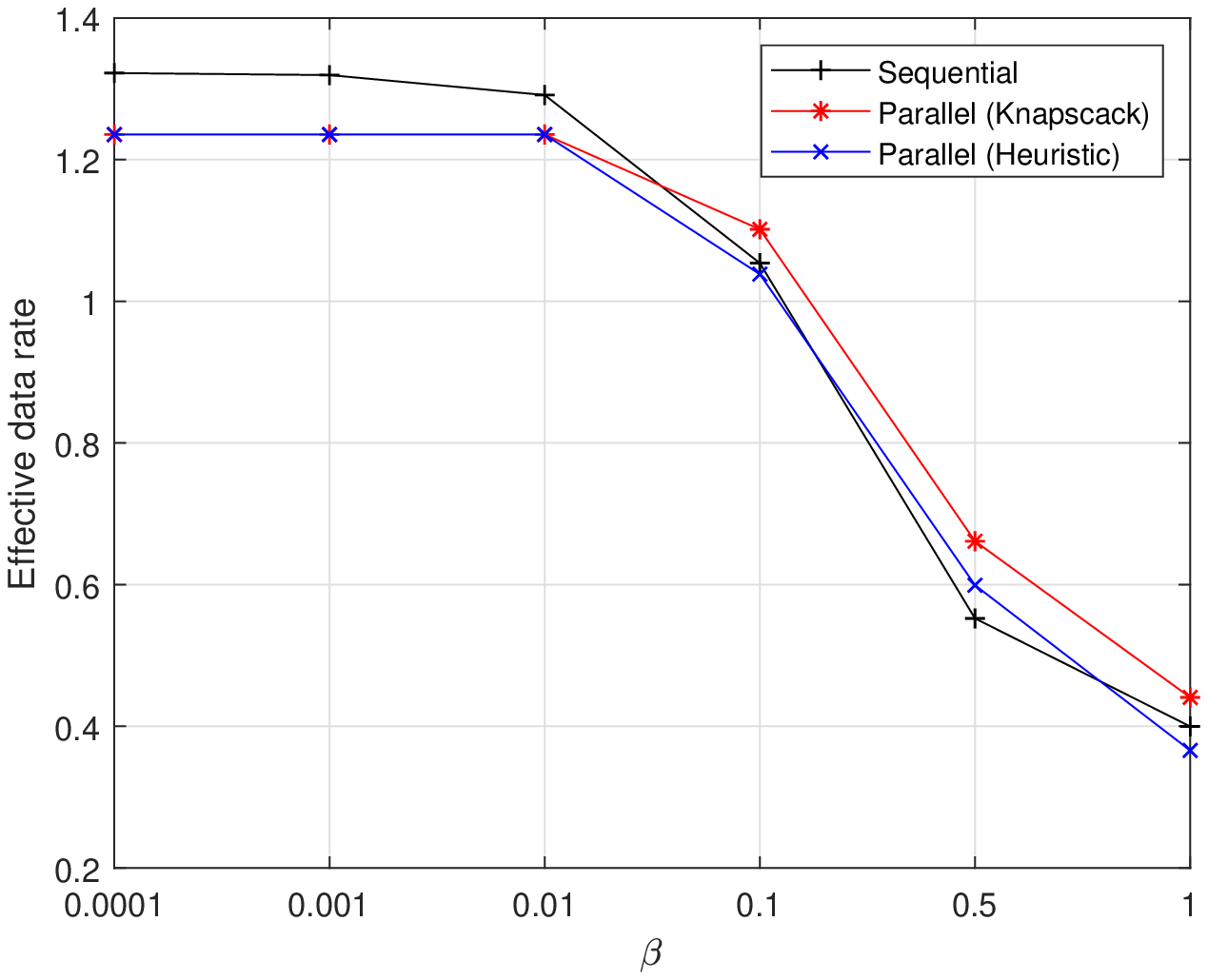}    
    \caption{\textbf{b)} Effective data rate achieved by parallel and sequential approaches when $N=12$, SNR$=5$dB, $\theta=100, \kappa=2$. } 
  \label{fig:n12_largetheta}
   \setcounter{figure}{7}
  \end{multicols}
  \end{figure*}
  
  \begin{figure*}[t]
  \begin{multicols}{2}
    \includegraphics[width=0.48\textwidth]{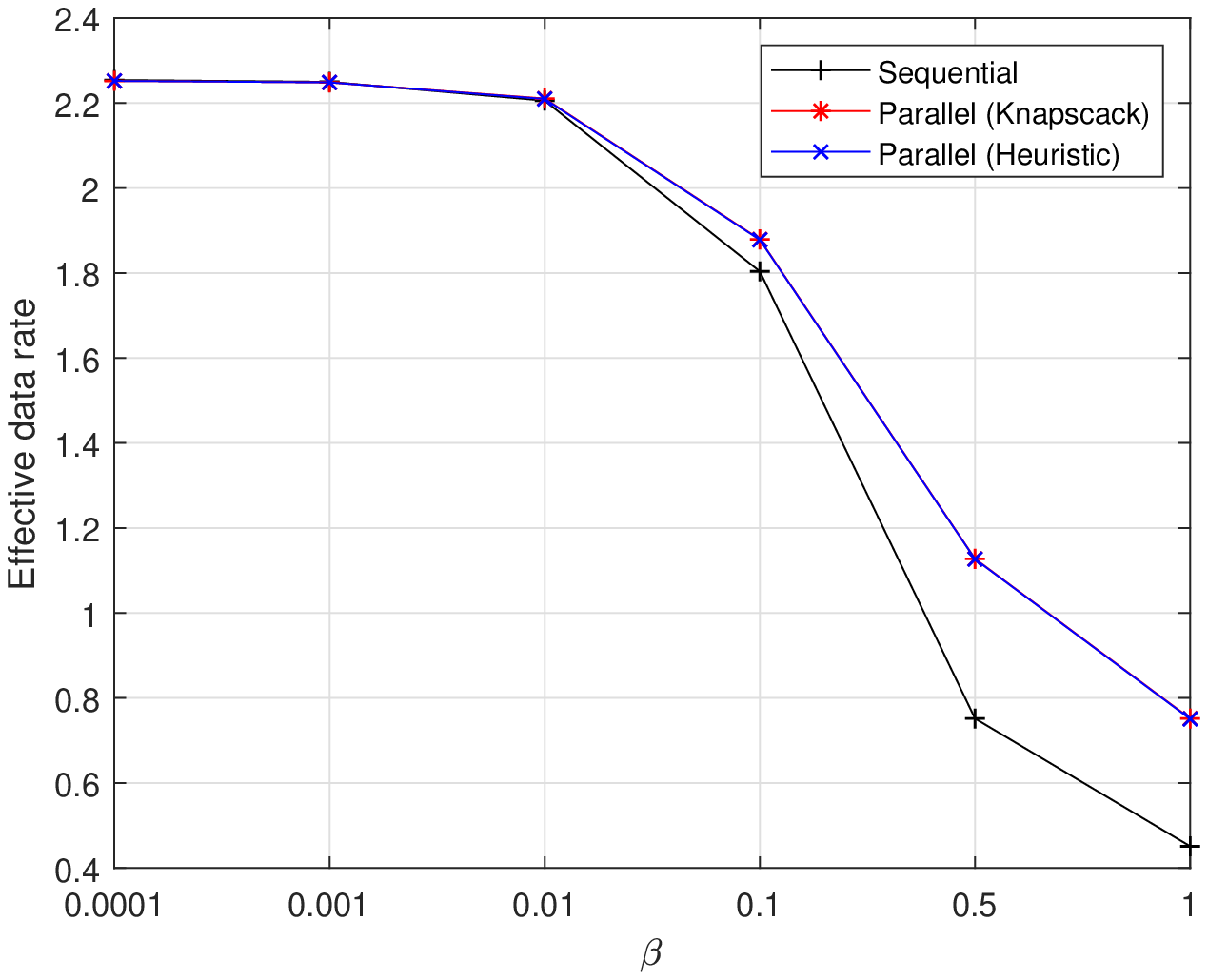}
     \caption{\textbf{c)} Effective data rate achieved by parallel and sequential approaches when $N=64$, SNR$=5$dB, $\theta=0.0001, \kappa=2$.} 
\label{fig:n64_smalltheta}
 \setcounter{figure}{7}
     \includegraphics[width=0.48\textwidth]{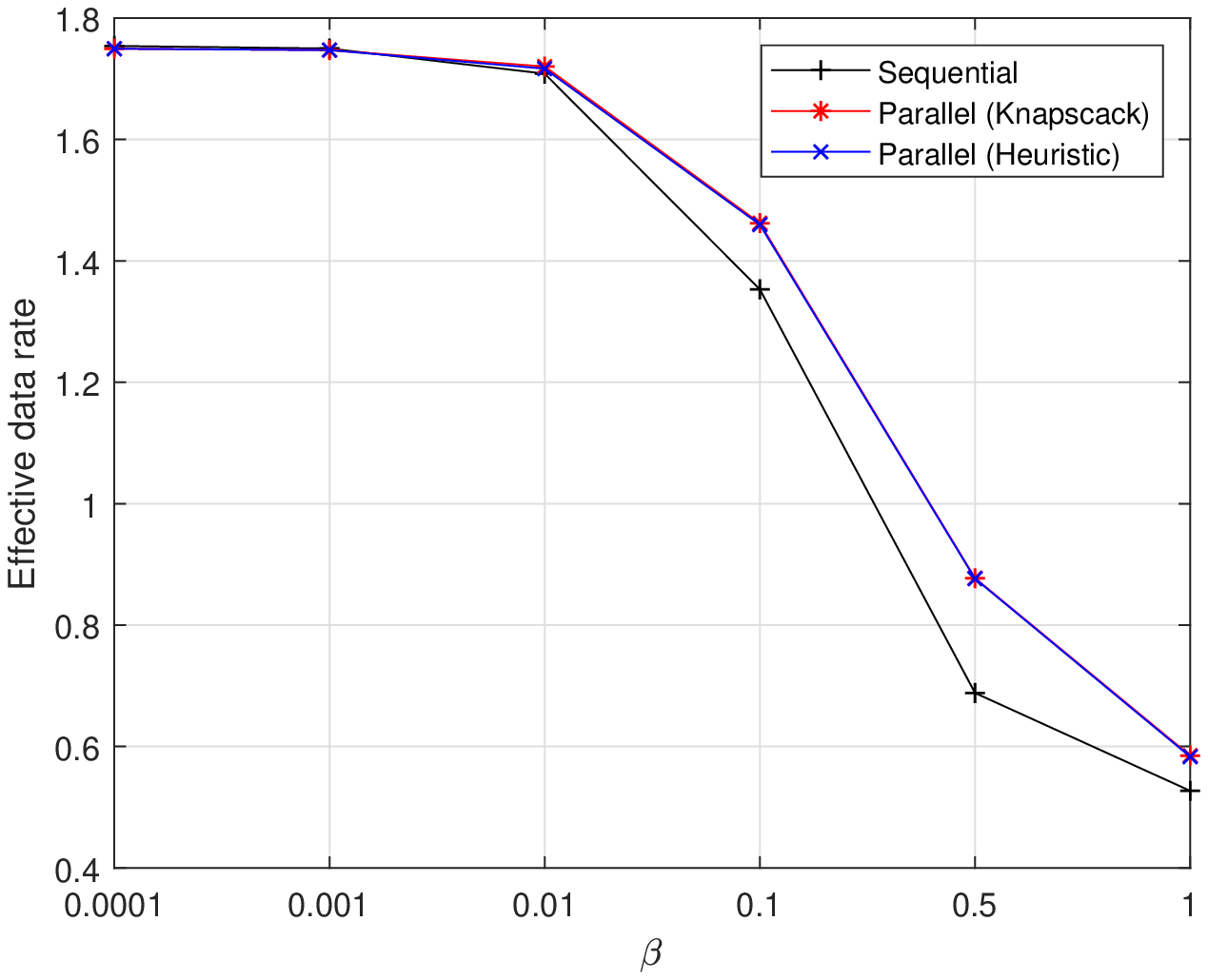}
     \caption{\textbf{d)} Effective data rate achieved by parallel and sequential approaches when $N=64$, SNR$=5$dB, $\theta=100, \kappa=2$.}
\label{fig:n64_largetheta}
   \end{multicols}
  \end{figure*}


\textcolor{black}{Now, we take a closer look and transform some specific cases from the 3d plots to two-dimensional graphs.} In Fig. \ref{fig:n12_smalltheta} we see the achieved \emph{effective data rate} $E_{C\mathcal,{D}}(\theta)$ given in \eqref{eq:eff_rate}, for different values of $N$ and $\theta$ while the SNR=$5$ dB and $\kappa=2$. Fig. \ref{fig:n12_smalltheta}a gives the achieved effective rate on set $\mathcal{D}$ for $N=12$ and $\theta=0.0001$ (relaxed delay constraint). Similarly to the case of long term average value of $C_D$ we see that for small values of $\beta$ the sequential approach achieves slightly higher effective data rate. As before, the increase of $\beta$ results in more reconciliation frames $M$ required in the sequential case. This effect is not seen in the parallel case and for high values of $\beta$ it performs better.

Fig. \ref{fig:n12_largetheta}b illustrates the case when $N=12$ and $\theta=100$ (very stringent delay constraint). \textcolor{black}{Similarly as in Fig. \ref{fig:surfn64}, we can see that for small values of $\beta$ the sequential approach performs better than the parallel. As discussed, the efficiency loss is caused by the fact that the devoted part of the total achievable effective capacity $E_{C}^{\text{opt}}(\theta)$ to reconciliation (syndrome communication) is more than what is required.} However, a higher $\beta$ leads to an increase in the reconciliation information that needs to be sent, and the rate of the subcarriers in set $\mathcal{\breve{D}}$ will be fully or almost fully utilised and the parallel approach shows better performance for these values.

In the next two Fig.: 
\ref{fig:n64_smalltheta}c and \ref{fig:n64_largetheta}d we show the performance of the two algorithms for higher value of $N=64$. It is easy to see that regardless of the value of $\theta$ and $\beta$ both algorithms perform identical or the parallel is better. In the previous case of $N=12$ increasing $\theta$ might reduce the effectiveness of the parallel approach, however when $N=64$ increasing $\theta$ does not incur such a penalty and the parallel is either identical to the sequential or outperforms it. 

Another interesting fact from Fig. \ref{fig:n12_smalltheta} is that looking at the parallel approach, it can easily be seen that in all cases the heuristic approach almost always performs as well as the optimal knapsack solution. The case of small values of $\theta$ is similar to the one when we work with long term average rate and choosing the best subcarriers for data transmission works as well as the optimal Knapsack solution. Interestingly, \textit{Algorithm 1} works well for high values of $\theta$, too. This can be explained by the fact that when $\theta$ increases the rate on all of the subcarriers becomes similar and switching the subcarriers in set $\mathcal{D}$ does not incur high penalty.





\section{Conclusions}
\label{sec:conclusions}
In this work we discussed the possibility of using SKG in conjunction with PUF authentication protocols, illustrating this can  greatly reduce the authentication and key generation latency compared to traditional mechanisms. Furthermore, we presented an AE scheme using SKG and a resumption protocol which further contribute to the system's security and latency reduction, respectively.

 In addition, we explored the possibility of pipelining encrypted data transfer and SKG in a Rayleigh BF-AWGN environment.  
We investigated the maximization of the data transfer rate in parallel to performing SKG. We took into account imperfect CSI measurements and the effect of order statistics on the channel variance. Two scenarios were differentiated in our study: i) the optimal data transfer rate was found under power and security constraints, represented by the system parameters $\beta$ and $\kappa$, which represent the minimum ratio of SKG rate to data rate and the maximum ratio of SKG rate to reconciliation rate; ii) by adding a delay constraint, represented by parameter $\theta$, to the security and power constraint we found the optimal \emph{effective data rate}. 

To finalise our study we illustrated through numerical comparisons the efficiency of the proposed parallel method, in which SKG and data transfer are inter-weaved to a sequential method where the two operations are done separately. The results of the two scenarios showed that in most of the cases the performance of both methods, parallel and sequential, is either equal or the parallel performs better. As the possible advantage of using the sequential is small and only applies in particular scenarios, we recommend the parallel scheme as a universal mechanism for general protocol design, when latency is an issue. Furthermore, a significant result is that although the optimal subcarrier scheduling is an NP hard $0-1$ knapsack problem, it can be solved in linear time using a simple heuristic algorithm with virtually no loss in performance.


\section*{List of abbreviations}\textcolor{black}{
\textbf{AE:} Authenticated encryption \\
\textbf{BF-AWGN:} Block fading additive white Gaussian noise \\
\textbf{B5G:} Beyond 5G \\
\textbf{CRP:} Challenge-response pair \\
\textbf{CSI:} Channel state information \\
\textbf{EAP-TLS:} Extensible authentication protocol-transport layer security  \\
\textbf{IoT:} Internet of things \\ 
\textbf{MAC:} Message authentication code \\
\textbf{MiM:} Man in the middle  \\
\textbf{NB-IoT:} Narrow band IoT\\
\textbf{OFDM:} Orthogonal frequency division multiplexing \\
\textbf{PHY:} Physical layer \\
\textbf{PKE:} Public key encryption \\ 
\textbf{PLS:} Physical layer security\\
\textbf{PUF:} Physical unclonable function\\ 
\textbf{QoS:} Quality of service\\
\textbf{RAN:} Radio access network \\
\textbf{RSS:} Received signal strength \\
\textbf{SKG:} Secret key generation\\
\textbf{SNR:} Signal-to-noise ratio\\
\textbf{STEK:} Session ticket encryption key \\
\textbf{TLS:} Transport layer security \\
\textbf{URLLC:} Ultra reliable low latency communication \\
\textbf{V2X:} Vehicle-to-everything communication\\
\textbf{0-RTT:} Zero round trip time \\
\textbf{3GPP:} The 3rd Generation Partnership Project}

\section*{Declarations}

\subsection*{\textit{Availability of data and material}}

No data sets were used in the production of the results shown in this paper. All the results can be regenerated from first principals using the formulations derived within the paper.

\subsection*{\textit{Competing interests}}

The authors declare that they have no competing interests.

\subsection{\textit{Funding}}

Miroslav Mitev is supported by the Doctoral Training Programme of CSEE, University of Essex, Arsenia Chorti is supported by the ELIOT ANR-18-CE40-0030 and FAPESP 2018/12579-7 and the INEX Project eNiGMA, Martin Reed  is supported by the project SerIoT which has received funding from the European Union’s Horizon 2020 Research and Innovation programme under grant agreement No 780139, Leila Musavian is supported by the RECENT project which has received funding from European Union Horizon 2020, RISE 2018 scheme (H2020-MSCA-RISE-2018) under Marie Sklodowska -- Curie grant agreement No. 823903.

\subsection{\textit{Authors' contributions}}

MM, AC, MR conceived this study. LM contributed on the use of Effective Capacity within this framework. MM carried out the simulations and prepared the graphs. All authors contributed and edited the manuscript. All authors read and approved the final manuscript.





\bibliographystyle{spmpsci}      
\bibliography{refs}

\end{document}